\documentclass[12pt,letterpaper, hidelinks]{article}

\usepackage{latexsym}
\usepackage{amssymb,amsmath, bm}
\usepackage{mathtools}
\usepackage{graphicx}
\graphicspath{{./figs/}}
\usepackage{marvosym}
\usepackage{multirow}
\usepackage{subcaption}
\usepackage{floatrow}
\newfloatcommand{capbtabbox}{table}[][\FBwidth]

\usepackage{comment}
\usepackage{booktabs}

\usepackage{tikz}
\usepackage{inputenc}
\usetikzlibrary{shapes,arrows,trees,fit,positioning,shapes.misc}

\usepackage[backend=biber,style=authoryear,autocite=inline,uniquename=false,natbib]{biblatex}
\renewbibmacro{in:}{%
  \ifentrytype{article}{}{\printtext{\bibstring{in}\intitlepunct}}}
\usepackage{dcolumn}
\newcolumntype{.}{D{.}{.}{-1}}
\newcolumntype{d}[1]{D{.}{.}{#1}}

\usepackage{theorem}
\theoremstyle{plain}
\theoremheaderfont{\scshape}
\newcommand{\qed}{\hfill \ensuremath{\Box}}

\usepackage{xr}

\usepackage{kantlipsum}
\allowdisplaybreaks

\usepackage{rotating}

\usepackage{arydshln}

\usepackage[compact]{titlesec}

\newcommand{\blind}{0}

\usepackage{times}

\usepackage[bookmarksopen=true, bookmarksnumbered=true,
pdfstartview=FitH, breaklinks=true, urlbordercolor={0 1 0}, citebordercolor={0 0 1}]{hyperref}

\begin{document}

\newcommand\const{\mathrm{const.}}
\newcommand*\diff{\mathop{}\!\mathrm{d}}
\newcommand*\expec{\mathop{}\mathbb{E}}
\newcommand\numberthis{\addtocounter{equation}{1}\tag{\theequation}}
\newcommand{\appropto}{\mathrel{\vcenter{
  \offinterlineskip\halign{\hfil$##$\cr
    \propto\cr\noalign{\kern2pt}\sim\cr\noalign{\kern-2pt}}}}}

\newcommand\dynMMSBM{\textsf{dynMMSBM}}
\newcommand\dist{\buildrel\rm d\over\sim}
\newcommand\ind{\stackrel{\rm indep.}{\sim}}
\newcommand\iid{\stackrel{\rm i.i.d.}{\sim}}
\newcommand\logit{{\rm logit}}
\renewcommand\r{\right}
\renewcommand\l{\left}
\newcommand\E{\mathbb{E}}
\newcommand\PP{\mathbb{P}}
\newcommand\V{\mathbb{V}}
\newcommand{\argmax}{\operatornamewithlimits{argmax}}

\newcommand\spacingset[1]{\renewcommand{\baselinestretch}%
  {#1}\small\normalsize}

\spacingset{1.1}

\newcommand{\tit}{Dynamic Stochastic Blockmodel Regression for
  Network Data: Application to International Militarized Conflicts}


\if0\blind

{\title{\tit\thanks{The methods described in this paper can be
     implemented via the open-source statistical software, {\sf
       NetMix}, available at
     \url{https://CRAN.R-project.org/package=NetMix}.}}

 \author{Santiago Olivella\thanks{Assistant Professor of Political Science, UNC-Chapel Hill. Email: \href{mailto:olivella@unc.edu}{\texttt{olivella@unc.edu}}} \quad \quad Tyler Pratt\thanks{Assistant Professor of Political Science, Yale University. Email: \href{mailto:tyler.pratt@yale.edu}{\texttt{tyler.pratt@yale.edu}}} \quad \quad Kosuke
   Imai\thanks{Professor, Department of Government and Department of Statistics, Harvard
     University.  1737 Cambridge Street, Institute for Quantitative
     Social Science, Cambridge 02138. Email:
     \href{mailto:imai@harvard.edu}{\texttt{imai@harvard.edu}}, URL:
     \href{https://imai.fas.harvard.edu/}{\tt https://imai.fas.harvard.edu/}}}

  \date{First Draft: July 12, 2018 \\ This Draft: \today}

\maketitle
}\fi

\if1\blind \title{\bf \tit} \maketitle
\fi

\pdfbookmark[1]{Title Page}{Title Page}

\thispagestyle{empty}
\setcounter{page}{0}

\begin{abstract}
  A primary goal of social science research is to understand how
  latent group memberships predict the dynamic process of network
  evolution. In the modeling of international militarized conflicts, for instance,
  scholars hypothesize that membership in geopolitical coalitions
  shapes the decision to engage in conflict.  Such theories explain
  the ways in which nodal and dyadic characteristics affect the
  evolution of conflict patterns over time via their effects on group
  memberships.  To aid the empirical testing of these arguments, we
  develop a dynamic model of network data by combining a hidden Markov
  model with a mixed-membership stochastic blockmodel that identifies
  latent groups underlying the network structure. Unlike existing
  models, we incorporate covariates that predict dynamic node memberships in
  latent groups as well as the direct formation of edges between
  dyads. While prior substantive research often assumes the decision
  to engage in international militarized conflict is independent
  across states and static over time, we demonstrate that conflict
  is driven by states' evolving membership in geopolitical
  blocs.  Changes in monadic covariates like democracy shift states
  between coalitions, generating heterogeneous effects on conflict
  over time and across states.  The proposed methodology, which relies
  on a variational approximation to a collapsed posterior distribution
  as well as stochastic optimization for scalability, is implemented
  through an open-source software package.
\end{abstract}
  \bigskip 
  

\newpage
\spacingset{1.83}

\section{Introduction}
\label{sec:intro}

Social scientists often posit theories about the effects of latent
groups of actors on relational outcomes of interest over time. For
example, international relations scholars have examined the so-called
``democratic peace'' hypothesis, which states that blocs of actors ---
defined by their democratic institutions --- rarely engage in wars
amongst themselves \citep[e.g.,][]{oneal1999kantian}. Others argue that militarized conflict is driven by state membership in geopolitical coalitions
that evolve over time \citep{farber1997common}. These
theories define latent groups of actors that underlie the
structures of social and political networks, and stipulate how the formation and
evolution of these groups give rise to various behaviors
\citep{lorrain:white:1971}.

To aid the empirical testing of these theories, we develop a dynamic
model of social networks that extends the mixed-membership stochastic
blockmodel \citep[MMSBM;][]{airoldi:etal:2008}.  The MMSBM is a
popular generalization of the stochastic blockmodel
\citep[SBM;][]{wong:wang:1987}, 
which is a factor analytic model for network data characterized by
latent groups of nodes \citep{hoff:2009}.  Unlike the SBM, the MMSBM
allows nodes to instantiate a variety of group memberships in their
interactions with other nodes. We extend the classical MMSBM in three
ways.  First, we allow memberships in latent groups to evolve over
time according to a hidden Markov process. Second, we define a
regression model for both latent memberships and observed ties,
incorporating both dyadic and nodal attributes to explain the
formation of groups.  This relaxes the strict assumption of stochastic
equivalence for members of the same groups. Finally, we apply
collapsed variational inference and improve computational scalability of the model.

Our approach, which we call \dynMMSBM, therefore frees applied
researchers from the need to resort to a commonly used two-step
procedure to evaluate theories, whereby memberships are first
estimated, and then regressed on covariates of interest
\citep[e.g.,][]{wasserman:faust:1994}.  Furthermore, the proposed
model allows for the prediction of group membership and future network
ties of previously unobserved nodes.  To facilitate the application of
our proposed model, we develop a fast Bayesian inference algorithm by
relying on a variational approximation to the collapsed posterior
\citep{teh:newman:welling:2007}, using stochastic gradient descent to
accommodate large-scale networks while retaining both theoretical
properties of the approximation and practical run times
\citep{hoffman:etal:2013, gopalan:blei:2013}. \if0\blind We offer an
open-source software \textsf{R} package, {\sf NetMix} (available on
CRAN) that implements the proposed methodology. \fi

We use the \dynMMSBM{} to conduct a dynamic analysis of
international conflicts among states over the last two centuries.
Political scientists have long sought to explain the causes of
interstate conflict and predict its outbreak.  In the study of the aforementioned democratic
peace hypothesis, a significant
body of evidence attests to the low rate of conflict among democratic
dyads \citep[e.g.,][]{maoz1993normative,
  oneal1999kantian,imai:lo:2021}.
Others argue that the relationship is spurious, driven by impermanent
geopolitical coalitions that generated common interests among
democracies \citep[e.g.,][]{farber1997common, gowa2011ballots}.
Analysts of the democratic peace typically want to account for these
underlying coalitions, and in particular ask whether democratic
political systems encourage states to enter the same geopolitical
blocs --- a question our model is designed to address.

\subsection*{Related Models}

Methodologically, our work extends the growing literature on dynamic
modeling of network data that exhibit some degree of stochastic
equivalence. In addition to the SBM, a variety of models are generally
available to accommodate such networks. For instance, the latent
position cluster model \citep{handcock:raftery:2007} and the recently
developed ego-ERGM \citep{salter:murphy:2015} incorporate equivalence
classes into the latent distance and the ERGM models,
respectively. Although the more flexible SBM (and all SBM-based
models, such as ours) can capture disassortative relationships that
these other models have a harder time accommodating, they all share
the highly restrictive assumption that nodes play a single role in all
their interactions.

Models like the overlapping/multiple-membership SBM
\citep{latouche:etal:2011, kim:leskovec:2013} or the
MMSBM \citep{airoldi:etal:2008} fully
address this issue by allowing nodes to belong to multiple equivalence
classes. Typically, however, these models are limited by the fact that
they assume independence of group memberships over time and across
nodes, as well as independence of dyads conditional on the
equivalence structure.  This makes it difficult to accommodate
networks that display both stochastic equivalence and some degree of
heterogeneity across nodes (e.g., networks that have very skewed degree
distributions).

Subsequent work therefore relaxes some of these independence
assumptions. For instance, \citet{sweet:thomas:junker:2014}
incorporates dyadic covariates into the MMSBM, thus allowing for
connectivity patterns that are not exclusively the result of the
stochastic equivalence structure. And \citet{white:murphy:2016}
incorporates node-specific attributes as predictors of the
mixed-membership vectors, thus eliminating the assumption that all
nodes in an equivalence class are exchangeable. Recent work by
\citet{yan:etal:2018} has shown that likelihood-based estimators of
these covariate effect parameters have desirable asymptotic
properties, lending further confidence in the validity of these
extensions. The proposed \dynMMSBM{} derives from these developments,
allowing for dyadic covariates at the edge-formation stage and for
nodal predictors of the mixed-membership vectors.

Even more attention has been devoted to relaxing the assumption of
independence of networks observed over time, resulting in important
advances to apply the MMSBM in dynamic network settings
\citep[e.g.][]{xing:fu:song:2010, ho:xing:2014,
  fan:cao:xu:2015}. As most social networks have a temporal dimension,
being able to model the dynamic evolution of relational outcomes is of
paramount importance to applied researchers. However, while these
models offer flexible approaches to accounting for temporal dynamics,
they often rely on continuous state space approaches like the Kalman
filter, making it difficult to periodize a network's historical
evolution.

Since researchers typically periodize history into distinct ``epochs''
to make sense of a phenomenon's evolution, more discrete approaches to
network dynamics would be better suited to the typical needs of social
scientists. Accordingly, the \dynMMSBM{} relies on a hidden Markov
process to capture the evolution of equivalence class-based network
formation. Furthermore, by assuming that the blockmodel itself (i.e.,
the matrix of edge propensities across and within latent classes)
remains constant over time --- so that only memberships into classes are
allowed to evolve --- we avoid the issues of identification raised by
\citet{matias:miele:2017} that affect some of the earlier dynamic
MMSBM specifications.

To the best of our knowledge, our model is the first to simultaneously
address the need to incorporate dyadic and nodal attributes as well as
the need to account for temporal dynamics, in an effort to develop a
model that can be readily employed in applied research.

\section{Challenges of Modeling the Interstate Conflict Network}
\label{sec:data}

The study of interstate conflict is of great interest to international
relations scholars and policy makers.  The ability to predict violent
political clashes has attracted a large literature on conflict
forecasting
\citep[e.g.,][]{schrodt1991prediction,beck2000improving,ward2013learning,hegre2017introduction}.  In addition, scholars have sought to
understand how specific political institutions, processes, and power
asymmetries affect war and peace among states
\citep[e.g.,][]{barbieri1996economic, oneal2006does, hegre2008gravitating}.

When analyzing conflict data, the most common methodological approach
is to assume conditional independence of state dyad-year observations
given some covariates within the generalized linear model framework
\citep[e.g.,][]{gleditsch1997peace,gartzke2007capitalist,dafoe2013democratic}.
However, there are reasons to believe
conflict patterns violate this conditional independence assumption.
For centuries, states have managed conflict through formal and
informal coalitions.  Alliances, for example, affect the probability
of conflict both among allied states and between allies and
non-allies.  Many militarized conflicts (most notably, the World Wars)
are \textit{multilateral} in nature: states do not decide to engage in
conflict as a series of disconnected dyads, but are drawn into war or
maintain peace as a result of their membership in preexisting, often
unobserved groups.

Recent analyses have turned to network models to relax this
conditional independence assumption.  \citet{maoz2006structural}, for
instance, use a measure of structural equivalence among dyads as a
covariate in the logistic regression.  In turn,
\citet{hoff2004modeling} employ random effects to explicitly model
network dependence in dyadic data, and \citet{ward2007disputes}
apply the latent space model developed by \citet{hoff2002latent} to
international conflict. Similarly, \citet{cranmer2011inferential}
propose and apply a longitudinal extension of the exponential random
graph model (ERGM) to conflict data. While we build on this emerging body of
scholarship that seeks to model complex dependencies in the conflict
network, our approach addresses several challenges faced by these existing network modeling
strategies.

First, and although existing approaches can capture higher order dependencies in conflict relations, they do
not directly model the evolving geopolitical coalitions that shape
patterns of conflict.  Such a model would more closely reflect the
theoretical mechanisms explaining why democracies form a distinct
community of states that have achieved a ``separate peace'' among
themselves.  This behavior may arise from the norms of compromise
prevalent in democratic societies \citep{maoz1993normative}, the
ability of democratic states to credibly signal their intentions
\citep{fearon1994domestic}, 
or the process by which democracies select into conflicts
\citep{Bdm2004selectorate}.
 
A second limitation of network analyses of international conflict is the need to restructure monadic covariates like
democracy to fit a dyadic analysis.  This problem has exacerbated a
debate in the democratic peace literature regarding the appropriate
dyadic specification of democracy \citep[see][]{dafoe2013democratic}.
An ideal model would directly incorporate nodal variables at the
country level by embedding them within the generative process of group
formation.  Finally, most existing methods do not provide flexibility
for the effect of democracy to vary over time, despite theoretical
claims that it should do so \citep{farber1997common,cederman2001back}.

In the following section, we propose a model that overcomes these
shortcomings.  The \dynMMSBM{} could uncover a democratic peace by
identifying a latent group that exhibits low rates of intra-group
conflict and that democratic states are more likely to join.  Other
hypotheses in this literature --- for example, the possibility of a
similar ``dictatorial peace'' among autocratic states
\citep{peceny2002dictatorial}, interactions between democracy and
power asymmetries \citep{Bdm2004selectorate}, and variation in the
strength of the democratic peace over time \citep{gleditsch1997peace,
  cederman2001back} --- are also accommodated by the model structure.
Each latent group is directly associated with its own set of nodal
covariates, and the dynamic implementation provides flexibility for
covariate effects to vary over time.
\section{The Proposed Model}
\label{sec:model}

Analyzing the interstate conflict network to study the democratic
peace theory requires a model that defines the probability of conflict
as a function of membership in latent groups of countries.  In
addition, the model must enable the exploration of how these
memberships evolve over time and how they are informed by
country-level characteristics --- particularly regime
type. Furthermore, for practical use, the model should deal with the
computational complexity involved in estimating a dynamic network
model with a large number of nodes.

Below, we describe a modeling approach that addresses these needs.  We
first define a general regression model for networked data, and then
derive a fast estimation algorithm based on a stochastic variational
approximation to the collapsed posterior distribution. While we focus
our exposition on directed networks, our model applies to undirected
networks with minimal modifications, as we illustrate in our
application.

\subsection{The Dynamic Mixed-Membership Stochastic Blockmodel}
\label{sec:modeldescription}

Let $G_t=(V_t, E_t)$ be a directed network observed at time $t$, with
node-set $V_t$ and edge-set $E_t$. For a pair of nodes $p,q \in V_t$,
let $Y_{pqt}=1$ if there exists a directed edge from node $p$ to $q$,
and $Y_{pqt}=0$ otherwise. Each node $i\in V_t$ is assumed to be
associated with a $K$-dimensional mixed-membership vector $\pi_{it}$,
encoding the extent to which $i$ belongs to each of $K$ latent groups
at time $t$.

To study how these mixed-memberships vary as a function of node-level
predictors, and to allow such memberships to evolve over time, we
further assume that the network at time $t$ is in one of $M$ latent
states, and that a Markov process governs transitions from one state
to the next. We then model each mixed-membership vector as a draw from
the following Markov-dependent mixture,
\begin{equation}
  \boldsymbol{\pi}_{it} \ \sim \ \sum_{m=1}^M  \Pr(S_t=m\mid S_{t-1})\times\text{Dirichlet}\l(\{\exp(\mathbf{x}_{it}^\top\boldsymbol{\beta}_{km})\}_{k=1}^K\r)
\end{equation}
where the vector of predictors $\mathbf{x}_{it}$ is allowed to vary
over time and the vector of coefficients $\boldsymbol{\beta}_{km}$ for
group $k$ is indexed by state $m$ in the Markov process.

Our model thus extends the MMSBM by allowing the mixed membership
vectors to not only be a function of node-level predictors, but also
by letting these vectors to change over time as the Markov states
evolve. Specifically, these random states are generated according to
$S_t\mid S_{t-1}=n \sim \text{Categorical}(\mathbf{A}_{n})$, which is
governed by a transition matrix $\mathbf{A}$ and the state at the
previous time period, $S_{t-1}$. We define a uniform prior over the
initial state $S_1$ and independent symmetric Dirichlet prior
distributions for the rows of $\mathbf{A}$.

The model is completed by defining a $K\times K$ blockmodel matrix
$\mathbf{B}$, with its $B_{gh}\in\mathbb{R}$ element giving
the propensity of a member of group $g$ to form a tie to a member of
group $h$ (for undirected network data, $\mathbf{B}$ is a
symmetric matrix).  Thus, we have,
\begin{align} Y_{pqt} & \ \sim \
\text{Bernoulli}\left(g^{-1}\l(\mathbf{z}^\top_{p\rightarrow
q,t}\mathbf{B}\mathbf{w}_{q\leftarrow p,t} + \mathbf{d}_{pqt}^\top
\boldsymbol{\gamma}\r)\right)
\end{align}
where $g^{-1}$ is the logistic function, and
$\mathbf{z}_{p\rightarrow q,t}\sim\text{Multinomial}(1, \boldsymbol{\pi}_{pt})$ is an
indicator vector for the group that node $p$ chooses when interacting
with node $q$ at time $t$ (and similarly for
$\mathbf{w}_{q\leftarrow p, t}$). To relax the assumption of strict
stochastic equivalence commonly used in other variants of the
stochastic blockmodel, we also incorporate dyadic predictors
$\mathbf{d}_{pqt}$ into the regression equation for the probability of
a tie, with regression coefficients $\boldsymbol{\gamma}$.

Put together, the data generating process can be summarized as follows:
\begin{enumerate}
\item For each time period $t>1$, draw a historical state $S_t\mid S_{t-1}=n \sim
\text{Categorical}(\mathbf{A}_{n})$.
\item For each node $i$ at time $t$, draw state-dependent mixed-membership vector \linebreak
  $\boldsymbol{\pi}_{it}\mid S_t=m \sim
  \text{Dirichlet}\l(\{\exp(\mathbf{x}_{it}^\top\boldsymbol{\beta}_{k,m}\}_{k=1}^K)\r)$.
\item For each pair of nodes $p$ and $q$ at time $t$,
  \subitem - Sample a group indicator $\mathbf{z}_{p \rightarrow q, t}\sim
  \text{Multinomial}(1, \boldsymbol{\pi}_{pt})$.
  \subitem - Sample a group indicator $\mathbf{w}_{q \leftarrow p, t} \sim
  \text{Multinomial}(1, \boldsymbol{\pi}_{qt})$.
  \subitem - Sample a link between them
  $Y_{pqt}  \ \sim  \
  \text{Bernoulli}\left(g^{-1}\l(\mathbf{z}^\top_{p\rightarrow
      q,t}\mathbf{B}\mathbf{w}_{q\leftarrow p,t} + \mathbf{d}_{pqt}^\top
    \boldsymbol{\gamma}\r)\right)$.
\end{enumerate}
This data generating process results in the following joint
distribution of observed and latent variables given a set of global
hyper-parameters $(\boldsymbol{\beta},\boldsymbol{\gamma},\mathbf{B})$
and covariates $(\mathbf{D}, \mathbf{X})$:
\begin{align}
  \begin{split}
     &P(\mathbf{Y},\mathbf{L},\boldsymbol{\Pi},
     \mathbf{A} \mid\boldsymbol{\beta}, \boldsymbol{\gamma},
     \mathbf{B},  \mathbf{D}, \mathbf{X})\\
    & =    P(S_{1})\l[\prod_{t=2}^T P(S_{t}\mid
    S_{t-1},\mathbf{A})\r] \l[\prod_{t=1}^T\prod_{it\in V_t}
    P(\boldsymbol{\pi}_{it}\mid \mathbf{X}, \boldsymbol{\beta}, S_t)\r]\prod_{m=1}^MP(\mathbf{A}_m)\\
    &\quad \times \l[\prod_{t=1}^T \prod_{p,q\in V_t} \Bigl[P(Y_{pqt}\mid \mathbf{z}_{p\rightarrow q,t},
    \mathbf{w}_{q\leftarrow p,t},\mathbf{B}, \boldsymbol{\gamma},
    \mathbf{D})   P(\mathbf{z}_{p\rightarrow
      q,t}\mid \boldsymbol{\pi}_{pt})P(\mathbf{w}_{q\leftarrow p,t}
    \mid \boldsymbol{\pi}_{qt})\r]
  \end{split}
\label{eq:joint}
\end{align}
where $\mathbf{L}:=\{\mathbf{Z},\mathbf{W},\mathbf{S}\}$ collects all latent group memberships and hidden Markov states,
$\boldsymbol{\Pi}:=\{\boldsymbol{\pi}_{it}\}_{it\in V_t}\forall t$ collects all mixed-membership vectors, and transition matrix $\mathbf{A}$ is defined as before.  

\subsection{Marginalization}
\label{subsec:marginal}

As we discuss in more detail in Section~\ref{subsec:estimation}, we
derive a factorized approximation to the posterior distribution
proportional to Equation~\eqref{eq:joint} in order to drastically
reduce the computation time required for inference. A typical
approximating distribution would factorize over all latent
variables. In the true posterior, however, latent group indicators
$\mathbf{z}_{p\rightarrow q,t}$ ($\mathbf{w}_{q \leftarrow p,t}$) and
the mixed-membership parameters $\boldsymbol{\pi}_{pt}$
($\boldsymbol{\pi}_{qt}$) are usually strongly correlated
\citep{teh:newman:welling:2007}.  Similarly, the Markov states $S_t$
and parameters in the transition kernel $\mathbf{A}$ are typically
highly correlated in the true posterior.

Therefore, and to avoid the strong assumption of independence induced
by the standard factorized approximating distribution, we marginalize
out the latent mixed-membership vectors and the Markov transition
probabilities and then approximate the marginalized posterior. The
details of the marginalization can be found in
Section~\ref{SI:collapse} of the Online Supplementary Information
(SI). Letting
$\alpha_{itkm}=\exp(\mathbf{x}_{it}^\top\boldsymbol{\beta}_{km})$,
$\alpha_{it\cdot m}=\sum_{k=1}^K\alpha_{itkm}$, and
$\theta_{pqtgh}=g^{-1}(B_{gh} +
\mathbf{d}_{pqt}^\top\boldsymbol{\gamma})$, the resulting collapsed
posterior is proportional to:
\begin{align}
 \begin{split}
      P(\mathbf{Y},\mathbf{L} &\mid
     \boldsymbol{\beta},\boldsymbol{\gamma},\mathbf{B},  \mathbf{X})\\ &\propto  \prod_{m=1}^M\left[
              \frac{\Gamma(M\eta)}{\Gamma(M\eta+U_{m\cdot})}\prod_{n=1}^M
              \frac{\Gamma(\eta+U_{mn})}{\Gamma(\eta)}\right]\\
            &\quad \times P(\mathbf{s}_1)
              \prod_{t=2}^T \prod_{m=1}^M \prod_{it\in V_t}
              \left[\frac{\Gamma(\alpha_{it\cdot m})}{\Gamma(\alpha_{it\cdot m}+
              2N_{t})} \prod_{k=1}^K \frac{\Gamma(\alpha_{itmk}+
              C_{itk})}{\Gamma(\alpha_{itmk})}\right]^{I(S_{t}=m)}\\
 &\quad  \times\prod_{t=1}^T  \prod_{p,q \in V_t} \prod_{g,h=1}^K
 \left(\theta_{pqtgh}^{y_{pqt}}(1-\theta_{pqtgh})^{1-y_{pqt}}\right)^{z_{p\rightarrow
     q,t,g}\times w_{q\leftarrow p,t,h}} \label{eq:coll}
   \end{split}
\end{align}
where $I(\cdot)$ is the binary indicator function, and $\Gamma(\cdot)$ is the
Gamma function.

The marginalized joint distribution explicitly use a number of
sufficient statistics:
$C_{itk} = \sum_{q \in V_t}(z_{i \rightarrow q,t,k} + w_{i\leftarrow
  q, t,k})$, which represent the number of times node $i$ instantiates
group $k$ across its interactions with all other nodes $q$ present at
time $t$ (whether as a sender or as a receiver);
$U_{mn}=\sum_{t=2}^{T}I(S_{t}=n)I(S_{t-1}=m)$, which counts the number
of times the hidden Markov process transitions from state $m$ to state
$n$; and $U_{m\cdot}=\sum_{t=2}^{T}\sum_{n}I(S_{t}=n)I(S_{t-1}=m)$,
which tracks the total number of times the Markov process transitions
from $m$ (potentially to stay at $m$).

\subsection{Estimation via Variational Expectation-Maximization}
\label{subsec:estimation}

For posterior inference, we rely on a mean-field variational
approximation to the collapsed posterior distribution
\citep{jordan:etal:1999, teh:newman:welling:2007}. To do so, we
define a factorized distribution over the latent variables
$\mathbf{L}$ as
\begin{equation} \widetilde{Q}( \mathbf{L}\mid
\mathbf{K},\boldsymbol{\Phi},\boldsymbol{\Psi}) \ = \ \prod_{t=1}^T
Q_1( \mathbf{s}_t \mid \boldsymbol{\kappa}_{t})\prod_{p,q \in V_t}
Q_2(\mathbf{z}_{p\rightarrow q,t} \mid
\boldsymbol{\phi}_{p\rightarrow q,t}) Q_2(\mathbf{w}_{q\leftarrow
p,t}\mid \boldsymbol{\psi}_{q\leftarrow p,t}),
\end{equation} where $\boldsymbol{\kappa}_{t}$,
$\boldsymbol{\phi}_{p\rightarrow q,t}$, and
$\boldsymbol{\psi}_{q\leftarrow p,t}$ are variational
parameters.  Our factorized approximation assumes
the latent state variables are independent in the collapsed
space. This is a strong assumption, but one that has been found to
strike a good balance between accuracy and scalability
\citep[see][]{wang:blunsom:2013}.

We then apply Jensen's inequality to derive a lower bound for the log
marginal probability of our network data $\mathbf{Y}$
\begin{equation}
 P(\mathbf{Y}\mid \boldsymbol{\beta},\boldsymbol{\gamma},\mathbf{B},\mathbf{X}) \geq \mathcal{L}\triangleq\expec_{\widetilde{Q}}[\log
P(\mathbf{Y},\mathbf{L}\mid\boldsymbol{\beta},\boldsymbol{\gamma},
\mathbf{B},\mathbf{X})] -\expec_{\widetilde{Q}}[\log
\widetilde{Q}(\mathbf{L}\mid
\mathbf{K},\boldsymbol{\Phi},\boldsymbol{\Psi})]
\label{eq:lb}
\end{equation}
and optimize this lower bound with respect to the variational
parameters to approximate the true posterior over our latent variables
\citep{jordan:etal:1999}. To do so, we iterate between finding an optimal $\widetilde{Q}$ (the
E-step) and optimizing the corresponding lower bound with respect to
the hyper-parameters $\mathbf{B}$, $\boldsymbol{\beta}$ and
$\boldsymbol{\gamma}$ (the M-step).

After initializing all sufficient statistics and variational
parameters, our E-step begins by updating the $\boldsymbol{\phi}$
parameters for all  $(pt,qt)$ dyads in our data as follows: 
\begin{align} \hat{\phi}^{(s)}_{p\rightarrow q, t,k} \ \propto\
\prod_{m=1}^M \left[\exp\bigl[\expec_{f,\widetilde{Q}_2}[\log
(\alpha_{ptmk}+C_{ptk}^\prime)]\bigr]\right]^{\kappa_{tm}}
\prod_{g=1}^K
\left(\theta_{pqtkg}^{y_{pqt}}(1-\theta_{pqtkg})^{1-y_{pqt}}\right)^{\psi_{q\leftarrow
p,t,g}}
\end{align}
where $C_{ptk}^\prime= C_{ptk}-z_{p\rightarrow q, t, k}$ and the
expectation is taken over the variational distribution of
$\mathbf{Z}$. By symmetry, the update for $\psi_{q\leftarrow p, t,k}$
is similarly defined.  In turn, and for $t=2, \ldots, T-1$, we update
all hidden Markov state variational parameters according to
\begin{align}
  \begin{split}
    \hat{\kappa}^{(s)}_{tm} & \ \propto \
\exp\left[-\!\expec_{\widetilde{Q}_1}[\log(M\eta +
U_{m\cdot}^\prime)]\right]
\exp\left[\kappa_{t+1,m}\kappa_{t-1,m}\expec_{f,\widetilde{Q}_1}[\log(\eta
+ U_{mm}^\prime + 1)]\right] \nonumber \\ &\quad
\times\exp\left[(\kappa_{t-1,m} - \kappa_{t-1,m}\kappa_{t+1,m} +
\kappa_{t+1,m})\expec_{f,\widetilde{Q}_1}[\log(\eta +
U_{mm}^\prime)]\right] \nonumber \\ &\quad \times \prod_{n\neq
m}\exp\left[\kappa_{t+1, n}\expec_{f,\widetilde{Q}_1}[\log(\eta +
U_{mn}^\prime)]\right] \prod_{n\neq m}\exp\left[\kappa_{t-1,
n}\expec_{f,\widetilde{Q}_1}[\log(\eta + U_{nm}^\prime)]\right]\\ &\quad
\times \prod_{pt\in V_{t}}
\left[\frac{\Gamma(\alpha_{it\cdot m})}{\Gamma(\alpha_{it\cdot m}+ 2N_{t})}
\prod_{k=1}^K \frac{\expec_{\widetilde{Q}_1}[\Gamma(\alpha_{ptmk}+
  C_{ptk})]}{\Gamma(\alpha_{ptmk})}\right] \nonumber,
\end{split}
\end{align}
where $U_{m\cdot}^\prime=U_{m\cdot}-s_{t,m}$ and
$U^\prime_{mn}=U_{mn}-s_{tm}s_{t+1,n}$. This definition of the term
$U^\prime_{mn}$ is valid whenever $m\neq n$ and $t\neq T$ (for other
cases, see Section~\ref{SI:VI} of the SI).

In order to avoid a costly computation of the Poisson-Binomial
probability mass function (which is required when computing
expected values that involve sufficient statistics), we approximate the
expectations in these updates by using a zeroth-order Taylor series
expansion, so that
$\expec_{f,\widetilde{Q}_2}[\log (\alpha_{ptkm}+C_{ptk}^\prime)] \approx
\log
\left(\alpha_{ptkm}+\expec_{f,\widetilde{Q}_2}\left[C_{ptk}^\prime\right]\right)$
and similarly for terms involving all $U^{\prime}_\cdot$ counts
\citep{asuncion:teh:2009}.

Finally, during the M-step, we find locally optimal values of
$\mathbf{B}$, $\boldsymbol{\beta}$ and $\boldsymbol{\gamma}$ with
respect to the following lower bound, using a quasi-Newton method (see
Section~\ref{SI:VI} of the SI for the expressions of the required
gradients),
\begin{align}
    \mathcal{L}_{\phi, \kappa}(\mathbf{B},\boldsymbol{\beta}, \boldsymbol{\gamma}) &\triangleq \sum^T_{t=1}\sum_{m=1}^M\kappa_{tm}\sum_{p\in V_t} \log\Gamma\left(\xi_{ptm}\right) - \log\Gamma\left(\xi_{ptm} + 2N_t\right) \nonumber\\
    &\quad + \sum^T_{t=1}\sum^M_{m=1}\kappa_{tm}\sum_{p\in V_t}\sum^K_{k=1}\expec_{\widetilde{Q}}[\log\Gamma(\alpha_{ptmk}+
    C_{ptk})] -
    \log\Gamma(\alpha_{ptmk})\nonumber\\
    &\quad +\sum^T_{t=1}\sum_{(p,q)\in E_{t}}
    \sum^K_{g,h=1} \phi_{p\rightarrow q,t,g}\psi_{q\leftarrow
      p,t,h}\l\{y_{pqt}\log (\theta_{pqtgh}) +(1-y_{pqt})\log(1-\theta_{pqtgh})\r\}\nonumber\\
    &\quad  - \sum^T_{t=1}\sum^M_{m=1}\sum_{(p,q)\in E_{t}} \sum_{k=1}^K \{\phi_{p\rightarrow
      q,t,k}\log( \phi_{p\rightarrow q,t,k})-
    \psi_{q\leftarrow
      p,t,h}\log(\psi_{q\leftarrow p,t,k}) \}
      \label{eq:lb_full}
\end{align}
To regularize the fit, we
define independent standard Normal priors for all parameters. When
required, standard errors for these quantities are obtained by first
sampling from the approximate posteriors of the latent variables, and
then obtaining expected values of the log-posterior Hessian evaluated
at the approximate MAP estimates of $\boldsymbol{\beta}$,
$\boldsymbol{\gamma}$, and $\mathbf{B}$.

\subsection{Stochastic VI Algorithm}
\label{subsec:svi}

For problems involving large networks, the above variational
approximation can be computationally intensive even after
parallelization (see Section~\ref{sec:impl}). To enable fast inference
on networks with a large number of nodes over multiple time periods,
we define an alternative optimization strategy which relies on the
stochastic gradient ascent approach proposed by
\cite{hoffman:etal:2013}, as applied to our collapsed variational
target \citep{foulds:etal:2013, dulac:etal:2020}.

Like other stochastic VI (SVI) algorithms, ours follows a random gradient
with expected value equal to the true gradient of the lower bound in
Equation~\ref{eq:lb}. To form this unbiased gradient, and at each step of the
algorithm, we sample a mini-batch of nodes within each time period $t$
uniformly at random, and form subgraphs $\mathbf{Y}^{(s)}_{t}$ among
all dyads containing the sampled nodes. The algorithm proceeds by
optimizing the local variational parameters (i.e. $\boldsymbol{\Phi}$
and $\mathbf{K}$) for all dyads $(p,q)$ in each $\mathbf{Y}^{(s)}_{t}$
using the updates given in the previous section, holding 
global counts constant at their most current values. We then condition on
these locally updated variational parameters and obtain an intermediate
value of all global counts (i.e. $\mathbf{C}$ and
$\mathbf{U}$) by computing their expected value under the mini-batch
sampling distribution.

We finalize each step by updating these global counts using a weighted
average:
\begin{equation}
    \mathbf{C}^{(s)} = (1-\rho_s)\mathbf{C}^{(s-1)}_{t} + \rho_s\expec_f\l[\mathbf{C}_{t}\r];\quad
    \mathbf{U}^{(s)}_t = (1-\rho_s)\mathbf{U}^{(s-1)} + \rho_s\expec_f\l[\mathbf{U}\r]
\end{equation}
where we set the step-size $\rho_s=(\tau+s)^{-p}$, and
$p\in (0.5, 1.0]$ and $\tau\geq0$ are researcher-set arguments
controlling the extent to which previous iterations affect current
values of the sufficient statistics
\citep{hoffman:etal:2013,cappe:moulines:2009}. To set the values of
our hyperparameters we once again follow an empirical Bayes approach,
updating the hyper-parameters along with the global sufficient
statistics by taking a step in the direction of the gradient of the
stochastic lower bound. As an example, for $\boldsymbol{\gamma}$, we
have:
\begin{equation}
  \boldsymbol{\gamma}^{(s)} =   \boldsymbol{\gamma}^{(s-1)} +
  \rho_s\nabla_{\gamma} \mathcal{L}^{(s)}_{\hat{\phi}, \hat{\kappa}}(\boldsymbol{\gamma})
\end{equation}
where
\begin{align*}
  \mathcal{L}^{(s)}_{\phi,\kappa}(\boldsymbol{\gamma})=\sum^T_{t=1}\frac{|E_t|}{|E^{(s)}_t|}\sum_{(p,q)\in E^{(s)}_{t}}
    \sum^K_{g,h=1} \phi_{p\rightarrow q,t,g}\psi_{q\leftarrow
  p,t,h}\bigl\{&y_{pqt}\log (\theta_{pqtgh})\\
  \quad &+(1-y_{pqt})\log(1-\theta_{pqtgh})\bigr\}
\end{align*}
is a random function that is equal to the third line in
Equation~\ref{eq:lb_full} in expectation. The updates for all other
hyper-parameters are similarly defined \citep{hoffman:etal:2013}.
Section~\ref{SI:VI} of the SI provides the required gradients.

When using the correct schedule for the step-sizes $\rho_s$, this
procedure is guaranteed to find a local optimum of the lower bound
without the need to perform a costly update over the parameters
associated with alls dyads at every iteration
\citep{gopalan:blei:2013}.

\subsection{Implementation Details}
\label{sec:impl}

Like other mixed-membership models, there are important practical
considerations when fitting the \dynMMSBM.  First, finding good
starting values is essential. In particular, the quality of starting
values for the sufficient statistics in the $\mathbf{C}$ global terms
proved to be highly consequential.  In our experience, two approaches
worked similarly well: an initial clustering based on a spectral
decomposition of the network's adjacency matrix \citep{Jin:etal:2018},
and taking a few samples from the posterior of the simpler
mixed-membership stochastic blockmodel (without covariates) of
\citet{airoldi:etal:2008}. We apply these strategies separately to
each time-stamped network, and resolve the ensuing label-switching
problem by re-aligning the (assumed constant) blockmodels using a
graph matching algorithm \citep{lyzinskietal2014}.

Second, and to establish convergence of our collapsed variational
algorithm, we evaluate absolute change in the estimated
hyper-parameters, and stop iterating when all changes fall below a
user-defined tolerance level ($10.0^{-4}$ in our application). In the
case of the SVI algorithm, we retain a small sample of dyads (viz. 1\%
of all pairs in our application) before initialization and evaluate
its log-likelihood after each iteration, stopping when average change falls
below a tolerance of $10.0^{-3}$ or when no improvement has been
observed in the past 20 iterations. The stopping rule based on a
held-out sample helps us avoid overfitting, and reduces the amount of
``jitter'' induced by the stochastic gradient descent. Finally, and to
maximize computational efficiency, we exploit the assumption of
conditional independence across edges and optimize local parameters
$\boldsymbol{\Phi}$ in parallel across (subsampled) dyads.

In Section~\ref{sec:simulation} of the SI, we conduct a series of
validation simulations, in which we evaluate the estimation accuracy
using a set of simulated dynamic networks, and compare the results of
fitting a fully specified \dynMMSBM{} and fitting a separate MMSBM
(without covariates) to each time period. We show the substantial
gains in error reduction resulting from the use of our proposed model.

\section{Empirical Analysis}
\label{sec:results}

We now apply the \dynMMSBM{} to study the onset of militarized
disputes among 216 states in the years 1816--2010, based on the
Militarized Interstate Dispute (MID) dataset version 4.1
\citep{MIDcite}.\footnote{The MID data are available at \url{https://correlatesofwar.org/data-sets/MIDs}}  The proposed model uncovers
the essential geopolitical coalitions that drive conflict patterns and
generates novel insights into the heterogeneous effect of key
covariates, like democracy.  Finally, we demonstrate that the
\dynMMSBM{} outperforms the standard logistic regression model in
forecasting future conflicts.

\subsection{The Setup}
\label{sec:conflictsetup}

We model conflict as an undirected network in which ties arise from
states' evolving membership in six latent groups.  While the
substantive results presented below are not sensitive to the number of
latent groups, we found that six provided sufficient flexibility to
model different types of evolving coalitions that can be qualitatively
interpreted.  Six latent groups also performed well in out-of-sample
prediction tests (see Table~\ref{tb:groups_AUC} in
Section~\ref{app:additional} of the SI for prediction tests and
Figures~\ref{fig:blockmodel5}~and~\ref{fig:blockmodel7} for a
visualization of blockmodel estimates for specifications with five and
seven groups).
 
A MID occurs when one state engages in a government-sanctioned
``threat, display or use of military force'' against ``the government,
official representatives, official forces, property, or territory of
another state'' \citep[168]{jones1996militarized}.  Ties in the
network are formed when a new dispute occurs between two states;
subsequent years of the same dispute are coded as 0.  The onset of a
MID is a rare event, only occurring in approximately $0.4\%$ of the
842,685 state dyad-year observations in our sample.

We include two node-level covariates $\mathbf{x}_{pt}$ --- the degree
of democracy in a state's domestic government and the state's military
capability --- that are hypothesized to influence membership in the
latent groups \citep{maoz1993normative, hegre2008gravitating}.  We
measure levels of democracy using the variable \verb|POLITY|, from the
Polity IV dataset \citep{marshall2017g}.  States are assigned a polity
score each year ranging from $-10$ to $10$, with higher values
representing more democratic political institutions.  The mean polity
score in our sample is $-0.43$.  Roughly six percent of state years
are assigned the minimum score of $-10$, and $16\%$ receive the
maximum of $10$.  Moreover, to measure the military capability of
states (\verb|MILITARY CAPABILITY|), we use version 5.0 of the
composite index \citep[CINC scores,][]{singer1972capability}, and take
the log to account for its skewed distribution.  The association
between these covariates and the latent group memberships is assumed
to depend on two hidden Markov states.

In addition, we include four dyadic variables $\mathbf{d}_{pqt}$ that
are expected to predict conflicts beyond the effects of the
equivalence classes induced by the blockmodel.  These include a
dichotomous indicator for a formal alliance between states in a given
year (\verb|ALLIANCE|); data on alliances comes from version~4.1 of
the COW Formal Alliances dataset \citep{Gibler2009}.  We also include
geographic distance (\verb|DISTANCE|) and the presence of a contiguous
border (\verb|BORDER|) between states
\citep{stinnett2002correlates}.\footnote{As an alternative way to
  address geographic effects, we estimate a specification that
  includes a set of regional indicator variables (see
  Table~\ref{tb:CWstates_region} and
  Figures~\ref{fig:blockmodel_regions}~and~\ref{fig:cluster_time_regions}
  in the SI).}  A count of common memberships in international
organizations (\verb|IO| \verb|CO-MEMBERSHIPS|) addresses the
possibility that interaction in these organizations decreases conflict
\citep{oneal1999kantian}.  Following the literature, we control for
further temporal trends using a count of years since the last
militarized dispute between each dyad and a cubic spline
\citep{beck:katz:tucker:1998}.  Finally, to account for the missing
values of some predictors, we rely on a missing-indicator approach,
adding dummy variables that indicate which observations have missing
values in the corresponding variable, and replacing all missing values
with zero.

The model is fitted using our open-source software package \if0\blind
{\sf NetMix}\fi.  Estimation took one hour and eighteen minutes on a
computer with a 3.6Ghz CPU, converging after 709 EM
iterations. Note that the estimation time drops to approximately 55
minutes without the optional Hessian computation, which calculates
standard errors for the blockmodel, monadic, and dyadic
coefficients.

\subsection{Memberships in the Latent Groups}
\label{sec:conflictmemberships}

\begin{figure}[t!] \spacingset{1}
  \centering
  \includegraphics[scale=.5]{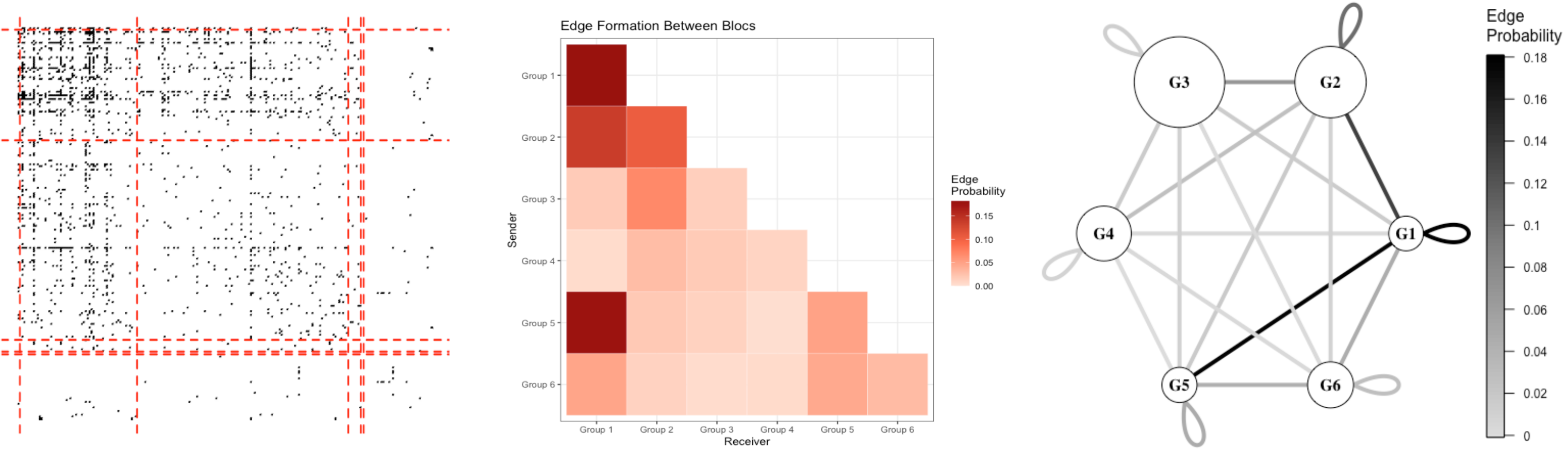}
  \caption{\textbf{Estimated blockmodel in the conflict network}. The
    left panel displays the adjacency matrix of militarized disputes
    between 216 states.  Black squares indicate the existence of at
    least one MID between the states in row \textit{x} and column
    \textit{y}; dotted lines separate states by estimated group
    membership.  The middle panel displays the estimated probability
    of conflict between members of groups as a heat map.  The right panel is a network graph summarizing the estimated blockmodel, where size of the nodes (circles) reflects aggregate membership in each group and weighted edges (lines) reflect the probability of conflict.}
  \label{fig:blockmodel}
\end{figure}

The \dynMMSBM{} allows us to characterize membership in each latent
group as well as the expected relationships between
them. Figure~\ref{fig:blockmodel} illustrates how patterns of
interstate conflict inform the estimation of group memberships.  The
left panel shows the 216 $\times$ 216 adjacency matrix of militarized
disputes between countries, aggregated over the entire time period.
Black squares indicate the existence of at least one MID between the
country represented by row \textit{x} and the country in column
\textit{y}.  The \dynMMSBM{} assigns each country to a mixture of the
six latent groups, each of which initiates disputes at unique rates.
In the matrix, we sort countries by estimated group membership --
demarcated in the figure by dotted lines -- to demonstrate the varying
rates of conflict within and between groups.

The middle panel of Figure~\ref{fig:blockmodel} shows the estimated
rates of conflict between groups.  For example, group~1 has elevated
rates of intra-group conflict as well as frequent conflict with
groups~2~and~5, as evidenced by the darker shade of these cells in the
figure.  Groups~4 and 5 have the most peaceful relations, initiating
disputes with each other 0.14\% of the time. Table~\ref{tb:blockmodel}
of Section~\ref{app:additional} in the SI presents the estimated
blockmodel used to create the figure.

The right panel combines information on group membership and dispute
rates, depicting each latent group as a node on a graph.  The size of
the nodes (circles) reflects the estimated membership size of the
group.  Group~3 is the most populous, representing 39.9\% of country-year
observations in the sample.  Group~2 is the second largest (27.2\%),
followed by Groups~4 (16.2\%), 6~(10.4\%), 5~(3.3\%),~and~1~(2.9\%).  The
edges (lines) depict the estimated rates of conflict between groups,
with darker-shaded edges indicating a higher propensity of conflict
onset.

To gauge the validity of these estimates, we examine whether the group
assignments and dispute probabilities correspond to known historical
conflict patterns.  Our model estimates that when a country from Group~1
interacts with a country from Group~2, there is an unusually high
probability (13.7\%) that a militarized dispute will occur between
them.  Probing the mixed-membership vectors of individual states
reveals that these two groups capture geopolitical divisions between
blocs of powerful states.  The United States, Canada, United Kingdom, 
and their Western European allies often instantiate Group~1, while
China, Russia, and other Eastern bloc countries tend to instantiate
membership in Group~2.

Other groups also reveal important structure in the international
system.  Group~3 includes many countries that maintained a foreign
policy of neutrality throughout much of the 19th and 20th centuries
(e.g., Norway, Finland, Ireland, and Costa Rica).  Despite their
neutral stance, these states maintained close diplomatic relations
with the Western allies that populate Group~1.  According to the
blockmodel, Group~3 has a low rate of conflict with Group~1~(1.7\%)
and is less bellicose overall.  Group~4 includes many countries that
were caught in the crossfire of the intense geopolitical conflict
between the Western and Eastern coalitions represented by
Groups~1~and~2.  Afghanistan, Angola, and Cambodia are among the
countries with high membership in Group~4 that were sites of proxy
conflicts during the Cold War period. Group~5 is composed of many
autocratic countries in the Middle East and Africa, while Group~6
features small or geographically remote countries.

A closer evaluation of estimated memberships during the Cold War era
lends further credence to the validity of the model.  As noted
earlier, this period was defined by a geopolitical rivalry between an
Eastern bloc, led by the Soviet Union, and a Western bloc, led by the
United States and its NATO allies.  To see if the \dynMMSBM{} recovers
the underlying geopolitical structure of the Cold War, we identify the
15 countries with the highest average membership probability in each
latent group during the period of 1950--1990.  We do this by computing
$ \frac{1}{T}\sum^{1990}_{t=1950} \pi_{ptg}$ for every country in a
given latent group $g$.  The countries with the highest membership in
each group are listed in Table~\ref{tb:CWstates} of
Section~\ref{app:additional} of the SI.

The group memberships of countries are consistent with presence of
competing geopolitical coalitions during the Cold War.  Group~1
contains the major NATO allies, including the United States, United
Kingdom, West Germany, Italy and Canada.  Non-NATO members that sided
with the NATO, including Japan and Australia, also instantiate Group~1
at high rates.  Group~2 consists of the Soviet Union and its allies in
the Eastern bloc (e.g.,, China, East Germany, Poland, Czechoslovakia,
and Romania).  The estimated blockmodel indicates the competing
coalitions experience abnormally high rates of conflict.

\subsection{The Dynamics of Membership}
\label{sec:conflictdynamics}

\begin{figure}[!t] \spacingset{1}
\includegraphics[scale=.6]{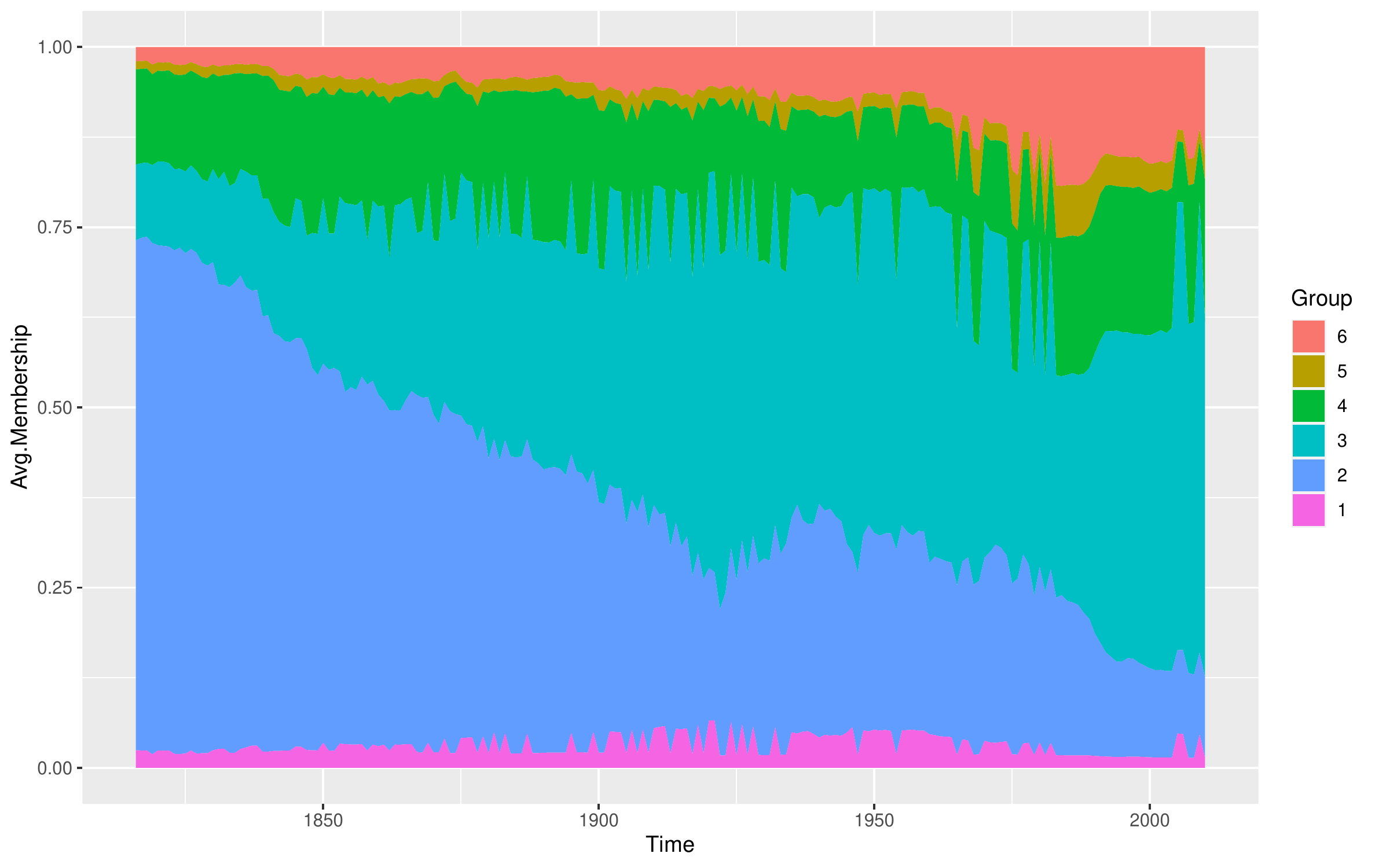} \vspace{.1in}
\caption[cluster time]{\textbf{Membership in Latent Groups over
Time}. The figure shows the average proportion of membership in six
latent groups for each year from 1816--2010.}\label{fig:cluster_time}
\end{figure}

The \dynMMSBM{} further allows us to examine how latent group
membership changes over time.  Figure~\ref{fig:cluster_time} displays
the evolution of group membership from 1816-2010.  Latent groups
expand and contract as countries move in and out of geopolitical
coalitions.  Group~2 --- populated by autocratic countries with high
military capacity --- noticeably declines in membership throughout the
period.  This reflects a general trend toward democratization among
industrialized countries, as well as geopolitical transitions of the
Soviet client states after the Cold War concluded.  The most peaceful
clusters, Group~3~and~4, increase in membership over the period, which
may be attributable to the consolidation of norms against military
aggression.  In the post-World War II era, decolonization and
independence movements led to a substantial increase in the number of
independent countries.  This likely accounts for the late growth of
Group~6 --- a cluster representing small countries with limited
military capability.

The evolution of groups shown in Figure~\ref{fig:cluster_time} are
consistent with international relations scholarship emphasizing
dynamic change in conflict patterns.  \citet{cederman2001back}, for
example, proposes a dynamic learning process in which democratic
countries consolidate peaceful relations over time.  The observed growth
of Group~3 --- a cluster populated by democracies with very low rates
of conflict --- supports this hypothesis.

\begin{figure}[!t] \spacingset{1} \centering
\includegraphics[scale=.55]{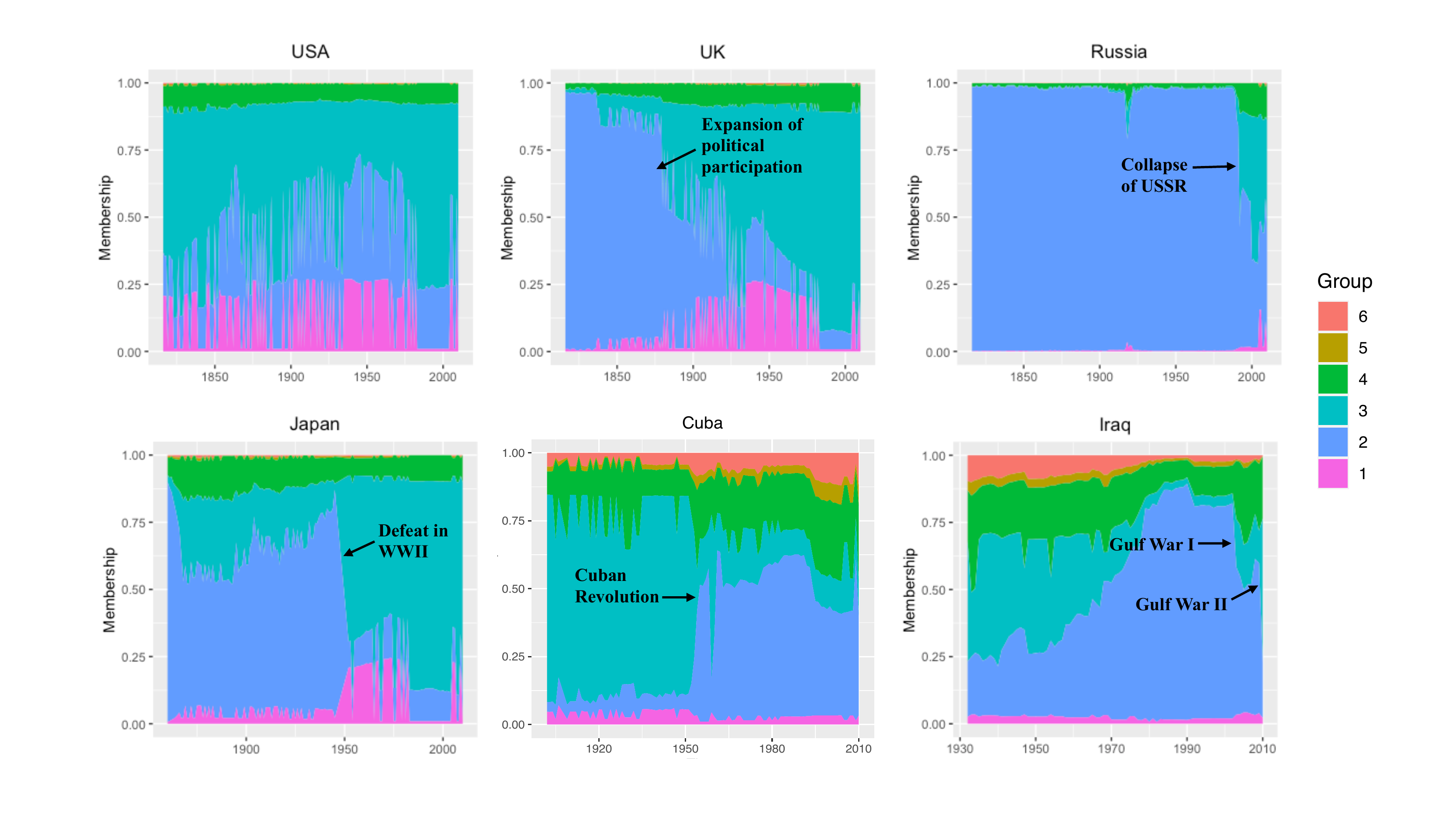} \vspace{-.1in}
\caption[cluster time node]{\textbf{Average Group Membership over Time,
Select Countries}. The figure shows, for six countries, the average rate of
membership in four latent groups in each year the country is present in
the network.}\label{fig:cluster_time_node}
\end{figure}

Figure~\ref{fig:cluster_time_node} displays the evolution of group
membership for a select group of countries.  There is significant
variation across countries and within some countries over time.  The
United States and United Kingdom feature relatively high membership in
Group~1 compared to other countries, as discussed above.  They also
exhibit significant membership in Group~3, the other Western-leaning
and democratic cluster.  US membership is comparably stable over the
period of the study, while the UK consolidates its membership in these
groups after transitioning to a democratic political system. For
example, we observe a sharp increase in the UK's membership in Group~3
following the 1867 Reform Act, which newly enfranchised parts of the
urban working class.  Russia's membership is overwhelmingly dominated
by Group~2.  At the end of the Cold War, the implosion of the Soviet
system shifts Russian membership toward Group~3 with a slight
reversion in the last few years.

Japan, Cuba, and Iraq further demonstrate how political shocks
like revolution and foreign intervention affect conflict patterns in
ways that are reflected in latent membership.  Japan experiences a
sudden shift from Group~2 to Groups~1 and~3 upon its defeat in World War II
and subsequent occupation by American forces.  The shift in membership
corresponds with a clear change in the country's conflict patterns.
Japan's overall rate of conflict declined from 2.7\% prior to 1945 to
0.7\% thereafter.  More than 60\% of Japan's disputes in the post-1945
period were with Group~2 members Russia, China, and North Korea.

Cuba's membership in Group~2 increases sharply following the onset of
the Cuban Revolution and the ascension of the Castro regime.  The
country experiences consistently high Group~2 membership since the
1950s, with a slight attenuation in the last few decades. In turn,
Iraq features two breaks in latent membership that correspond to
conflicts with the United States.  Following the first Gulf War in
1990-1991, we observe reduced membership in Group~2 and increases in
Groups~3 and ~4.  A similar shift in 2003 reflects the invasion by the
US and allied countries and the installation of a new
government.

\subsection{Covariate Effects}
\label{sec:cov_fx}

\begin{table}[t] \spacingset{1} \setlength{\tabcolsep}{6pt} \centering
{\footnotesize
  \begin{tabular}{llllllll}  \toprule \textbf{Predictor} &
\multicolumn{1}{c}{\textbf{Dyadic}} & \multicolumn{1}{c}{\textbf{Group
1}} & \multicolumn{1}{c}{\textbf{Group 2}} &
\multicolumn{1}{c}{\textbf{Group 3}} &
\multicolumn{1}{c}{\textbf{Group 4}} &
\multicolumn{1}{c}{\textbf{Group 5}} &
\multicolumn{1}{c}{\textbf{Group 6}} \\ \midrule {\tt INTERCEPT} & &
12.016 & 16.539 & 11.383 & 12.376 & 8.836 & 7.389 \\ & & (1.069) &
(1.069) & (1.069) & (1.069) & (1.074) & (1.066) \vspace{2mm} \\ {\tt
POLITY} & & 0.083 & -0.251 & 0.076 & -0.115 & -0.091 & -0.091 \\ & &
(1.084) & (1.083) & (1.084) & (1.083) & (1.096) & (1.079) \vspace{2mm}
\\ {\tt MILITARY} & & 0.638 & 1.192 & 0.130 & 0.513 & 0.235 &
-0.134 \\ {\tt CAPABILITY}& & (1.029) & (1.029) & (1.025) & (1.029) &
(1.048) & (1.059) \\ \midrule {\tt BORDERS} & 2.123 &&&&&& \\ &
(0.001) &&&&&&\vspace{2mm} \\ {\tt DISTANCE} & -0.0001 &&&&&& \\ &
(0.002) &&&&&& \vspace{2mm} \\ {\tt ALLIANCE} & 0.087 &&&&&& \\ &
(0.001)&&&&&& \vspace{2mm} \\ {\tt IO CO-MEMBERS} & 0.009 &&&&&& \\ &
(0.002) &&&&&& \vspace{2mm} \\ {\tt PEACE YRS} & -0.021 &&&&&& \\ &
(0.002) &&&&&&\\ \midrule \multicolumn{8}{l}{$N$ nodes: 216; $N$
dyad-years: $842,685$; $N$ time periods: 195}\\ \multicolumn{8}{l}{Lower
bound at convergence: $-527,587.7$}\\ \bottomrule
\end{tabular} }
\caption{\textbf{Estimated Coefficients and their Standard Errors}.
  The table shows the estimated coefficients (and standard errors) of
  the two monadic predictors for each of six latent groups, as well as
  those of the dyadic predictors for edge formation.  We present the
  results from the first Markov state, which accounts for the majority
  of the time period.  The estimated coefficients for cubic splines
  and indicators for variable missingness are not
  shown.}\label{tb:covFX}
\end{table}

The \dynMMSBM{} also enables the examination of covariate relations
that can help characterize the nature of each estimated latent group.
The upper panel of Table~\ref{tb:covFX} displays coefficient estimates
for the monadic covariates {\tt POLITY} and {\tt MILITARY CAPABILITY}.
The estimates represent the effect of each covariate on the log-odds
of membership in each latent group. In the interest of space, and
since the majority of the time period under study (viz. $51.3\%$) is
estimated to derive from this state, we display the coefficients only
for Markov state 1. See Table~\ref{tb:covFX_Markov2} in
Section~\ref{app:additional} of the SI for Markov state 2
coefficients.

Democratic regimes (i.e., those with high {\tt POLITY} scores) are
most likely to instantiate membership in Groups~1 and 3.  This is
consistent with the interpretation of Group~1 as the Western alliance
of liberal democracies during the Cold War, and Group~3 as
Western-leaning neutral states.  Notably, these two democratic
clusters exhibit different patterns of conflict.  Group~1 countries
have a high rate of military disputes, both with other Group~1 members
(18.2\%) and with other groups (7.7\%).  Group~3 countries are more
consistent with the democratic peace hypothesis.  Predicted conflict
between members of this group are rare (0.14\%), and they also have a
lower dispute rate with other latent groups (2.3\%).

Other monadic coefficients are largely consistent with the descriptive
patterns discussed above.  Autocratic regimes sort into Group~2 at the
highest rate.  Greater military capability is negatively associated
with membership in Group~6 and positively associated with membership
in the other clusters.

In addition to obtaining estimates for the coefficients in our model,
we can also predict how the probability of conflict changes as a
function of the node's monadic covariates.  In the generative model,
group memberships are instantiated for each dyad in each time period.
As a result, countries in the conflict network are assigned to a
latent group each time they interact with another country in a given
year.  Because the probability of edge formation depends on the group
membership of both nodes in a dyad, a change in one node's monadic
predictor will yield heterogeneous effects across dyads, nodes, and
time.

For example, consider the change in predicted conflict propensity when
each country's {\tt POLITY} score is increased by one standard deviation
(6.78), making sure scores
increase only up to the maximum value (10). The overall average effect of this
change on the probability of edge formation, averaging all dyadic
interactions and time periods,
\[ \frac{1}{T} \sum\limits^T_{t=1} \frac{1}{|V_t\times V_t|}\sum_{p,q
\in V_t} [ \expec(y_{pqt} \mid \verb|POLITY| + 6.78) - \expec(y_{pqt})
]
\] is negative but negligible in size: $-0.001$.  Thus, increasing the
degree of democracy in a country results in a minor decrease in
overall conflict, given the underlying geopolitical coalitions
throughout the time period.

\begin{figure}[!t] \spacingset{1}
  \begin{center} \includegraphics[scale=.75]{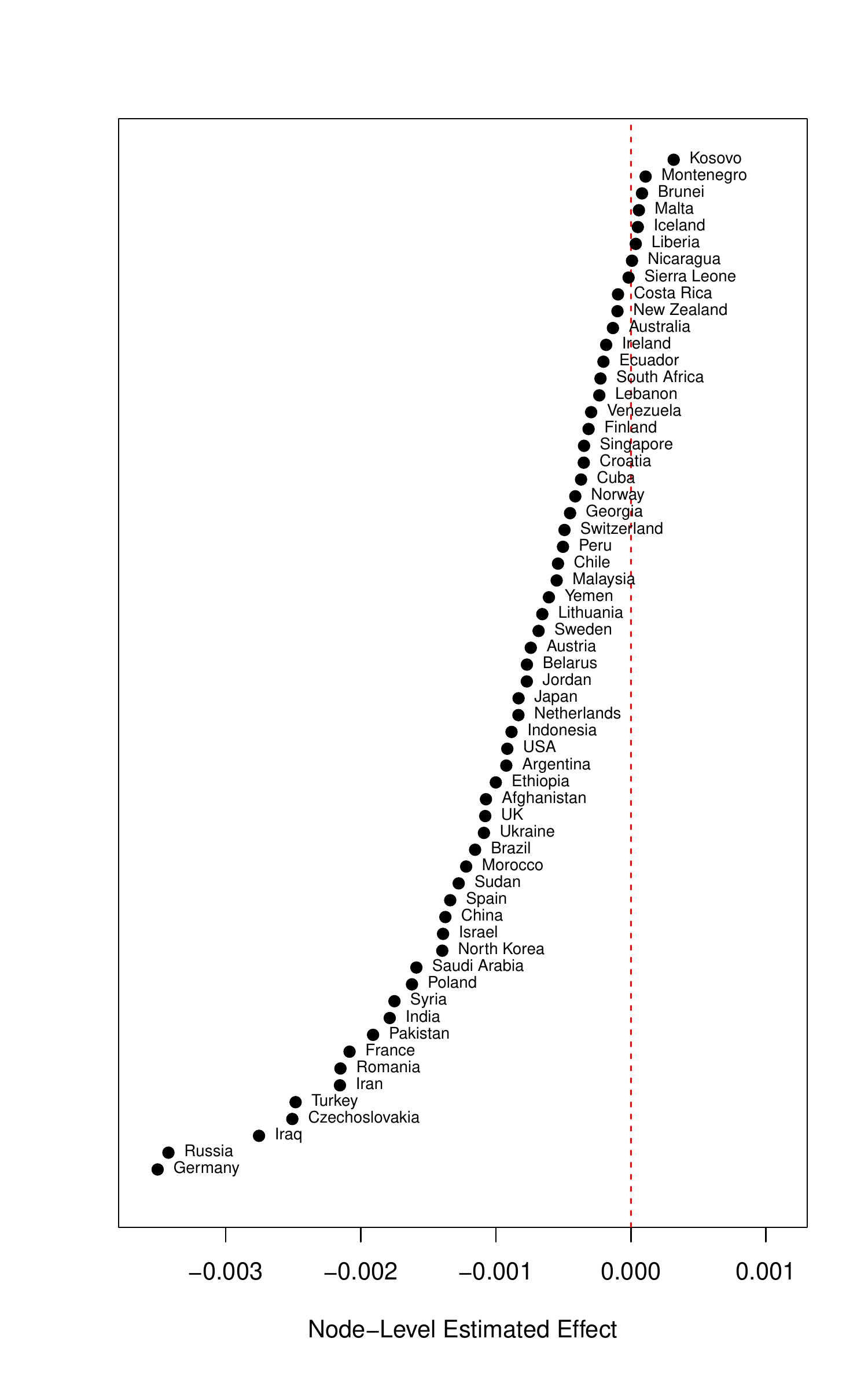}
 \end{center} \vspace{-.36in}
 \caption[CountryHet]{\textbf{Estimated Effects of Covariate Shift in
Polity over Time, Select States}.  The figure shows the estimated
change in the probability of interstate conflict if a state's {\tt
POLITY} score is increased by one standard deviation (6.78) from its
observed value.}\label{fig:CountryHet}
\end{figure}

There is, however, a significant amount of heterogeneity in this
effect across countries and over time.  Figure~\ref{fig:CountryHet}
shows, for a large set of countries, the difference in expected probability of
interstate conflict due to an increase of one standard deviation in
{\tt POLITY} score.  Many countries (such as Germany, Russia, and
Iraq) are predicted to be substantially more peaceful, on average, if
they were more democratic during the period of the study.  Others,
however, experience very little change in conflict behavior (e.g.,
Australia and Nicaragua).  A handful of countries are estimated to
become \textit{more} conflict prone (e.g., Kosovo, Montenegro, and
Brunei).  An increase in polity shifts these countries into different
latent groups that are more conflictual, on average.

\begin{figure}[!t] \spacingset{1}
  \begin{center} \includegraphics[scale=.615]{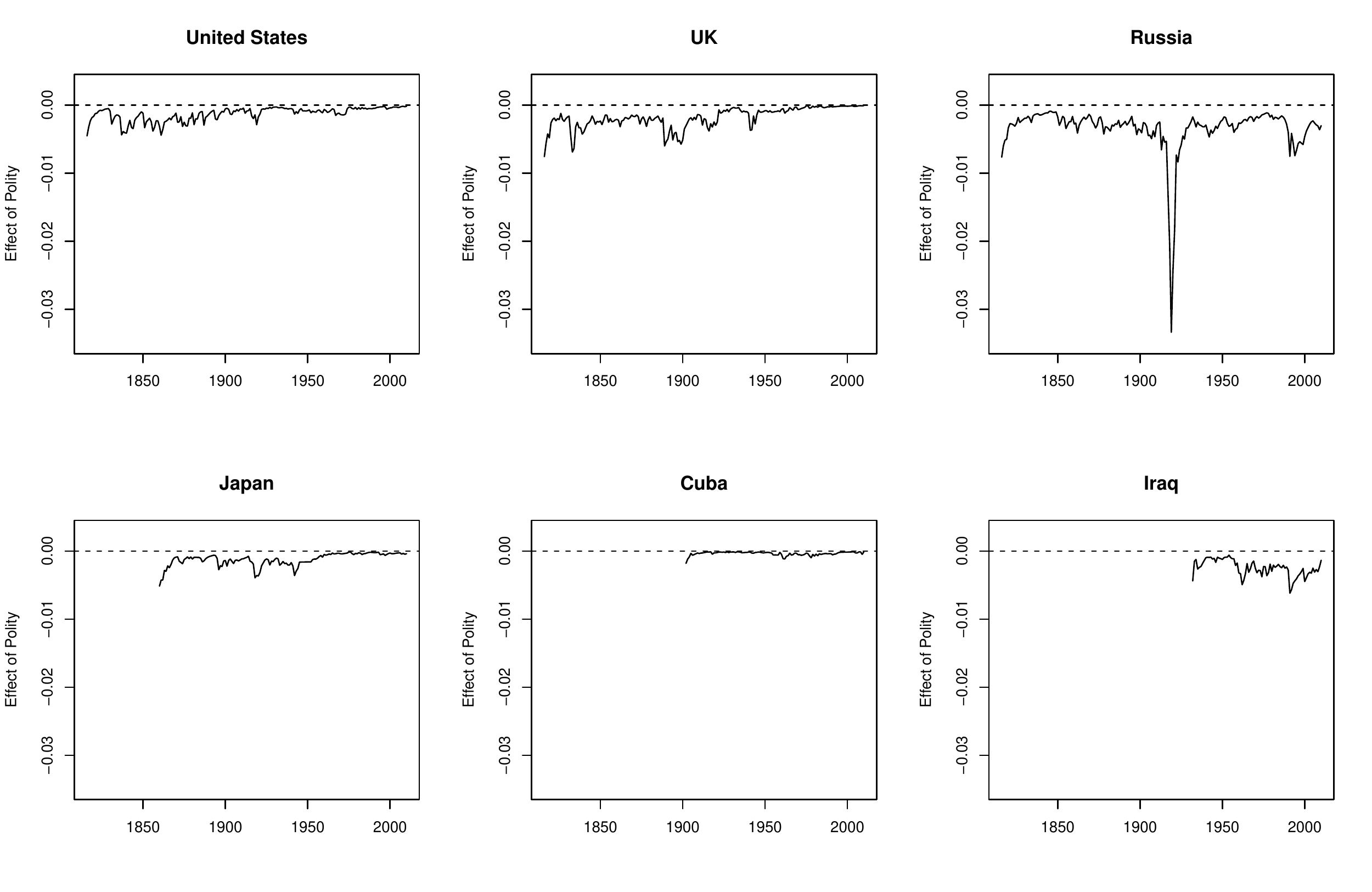}
 \end{center} \vspace{-.3in}
 \caption[CountryHet2]{\textbf{Effect of Shift in Polity over Time,
Select States}.  The figure shows the estimated change in the
probability of interstate conflict over time if a country's {\tt POLITY}
score is increased by one standard deviation (6.78) from its observed
value (up to a maximum of 10).}\label{fig:CountryHet2}
\end{figure}

\begin{figure}[!t] \spacingset{1}
  \begin{center} \includegraphics[scale=.65]{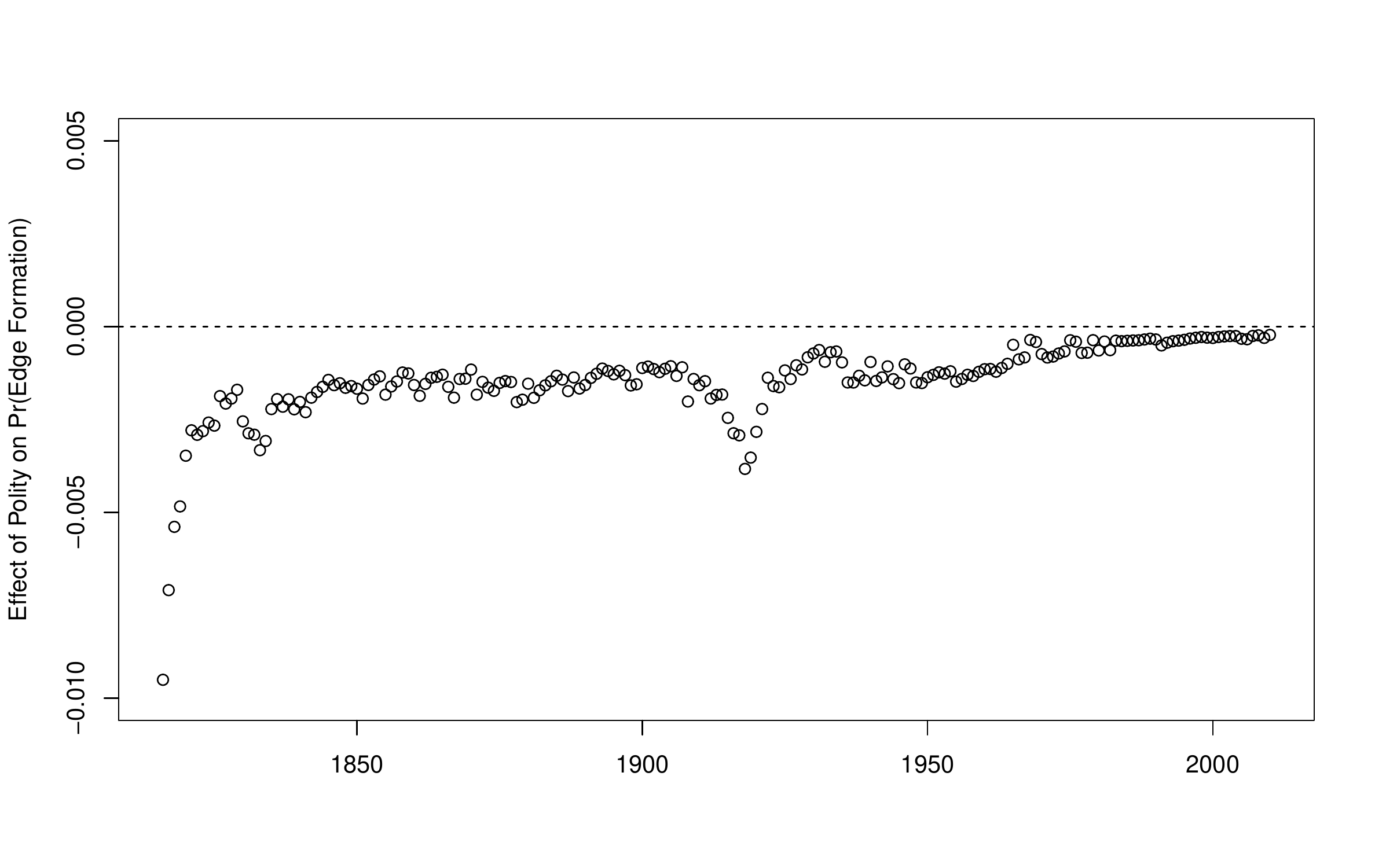}
 \end{center} \vspace{-.5in}
 \caption[politytime]{\textbf{Estimated Aggregate Effect of Shift in
Polity over Time}.  The figure shows the estimated average change in
the probability of interstate conflict when countries' {\tt POLITY}
scores are increased by one standard deviation (6.78) up to the maximum {\tt POLITY} score.}\label{fig:politytime}
\end{figure}

The effect of democracy varies due to the latent group structure of
the model.  In general, shifts in monadic predictors will generate
effects that are non-linear and contingent upon the existing group
membership of the node in question and the other nodes in the network.
Figure~\ref{fig:CountryHet2} looks within countries to gauge the
effect of the shift in \verb|POLITY| over time, revealing additional
heterogeneity.  To illustrate how monadic effects can vary within
countries, consider the sharp drop in the estimated effect of {\tt
  POLITY} for Russia from 1918-1921.  This period is preceded by the
ascendance of the Bolshevik government, which took power in November
1917.  Over the next few years, the government engaged in a series of
militarized disputes with the Allied Powers of WWI, who supported
anti-communist forces during the Russian Civil War.  This pattern of
disputes is consistent with the estimated blockmodel, which predicts
an elevated rate of conflict between Group~1 (US, UK, France, Japan)
and Group~2 (Russia).  The estimates in Figure~\ref{fig:CountryHet2}
compare these patterns of conflict to a counterfactual world in which
Russia had a more democratic political system.  Increasing Russia's
{\tt POLITY} score from its observed value in 1918 ($-1$) to a higher
value ($6$) shifts the expected group membership for Russia away from
Group~2 (from 75.4\% to 32.4\%) and toward Group~3 (from 10.3\% to
40.7\%).  This reduces the likelihood of disputes, since Group~3 has
significantly lower rates of inter- and intra-group conflict.  By 1922
the Bolshevik regime consolidated power and the country's {\tt POLITY}
score drops to $-7$, after which an equivalent increase in {\tt
  POLITY} has a smaller effect.

Figure~\ref{fig:politytime} displays the average effect of
\verb|POLITY| for each year in the time period.  An increase in
democracy induces less conflict, on average, throughout most of the
sample.  The effect is noticeably lower during the pre-WWII period,
hitting a local minimum in 1918 ($-0.004$).  The impact of polity has
attenuated in recent years, when the estimated effect of increasing
polity approaches zero.\footnote{To ensure these patterns are not a
  function of ceiling effects --- given that the number of states with
  the maximum polity score of 10 is increasing over the time period
  --- we also calculate the effect of a one standard deviation
  \textit{decrease} in polity (see
  Figures~\ref{fig:CountryHet2_decrease} and
  ~\ref{fig:politytime_decrease} in the SI). The effects are
  substantively identical.}

Finally, dyadic predictors operate outside the latent group membership
structure, directly influencing the probability of conflict among
states. In a sense, they serve as controls for alternative networks
defined on the same node set.  The dyadic coefficient estimates appear
in the bottom panel of Table~\ref{tb:covFX}.  Consistent with existing
work, sharing a border significantly increases the likelihood of
conflict.  Greater geographic distance between states has no
statistically discernible effect on conflict propensity.  Somewhat
surprisingly, the presence of a formal alliance and joint membership
in international organizations increase the likelihood of conflict,
though these effects are substantively small.

\subsection{Additional Analyses}

In Section~\ref{app:logitcompare} of the SI, we compare the results of
our empirical analysis with those of the standard logistic regression
model, which assumes all dyad-years are conditionally independent and
forces all node-level predictors to be transformed into dyadic
form. We furthermore emulate the process of analyzing data in
real-time by estimating both models using data from 1816-2008 and then
evaluating model performance on what the forecasting predictions would
have been during the two following years, 2009 and 2010.  We find that
the \dynMMSBM{} significantly outperforms the conventional approach in
the Diebold-Mariano test for forecasting comparison
\citep{Diebold1995}.  It also marginally improves on the logistic
model in area under the ROC curve, though the difference is not
statistically significant.

Our primary results reflect a batch analysis of the data, taking all
years into consideration.  In Section~\ref{app:logitcompare} of the
SI, we use out-of-sample prediction to evaluate forecasting
performance.  In Section~\ref{app:additional} of the SI, we replicate
our analysis via ``online'' updating, where we iteratively expand the
time window to update estimates as if the data had been obtained
sequentially, rather than in batch. To illustrate this approach, we
first fit a model for the years 1816--1820, then use the resulting
mixed membership estimates as starting values for a model that
incorporates the next window (1821-1825).  We repeat until all years
are included (see Table~\ref{tb:covFX_online} and
Figures~\ref{fig:cluster_time_online}~and~\ref{fig:cluster_time_node_online}).

\section{Conclusion}

We have introduced the \dynMMSBM, a generalization of the
mixed-membership stochastic blockmodel that incorporates dyadic and
nodal attributes, and accounts for episodic temporal evolution of
networks using a hidden-Markov process. The proposed model enables
researchers to evaluate dynamic theories about the role of individual
characteristics on the generation of relational outcomes when abstract
groups of actors are the driving force behind tie formations.  The
\dynMMSBM{} also helps identify periods in time when a network
exhibits distinctive patterns of interactions among actors.

Using a network defined by almost 200 years of militarized interstate
disputes in the international system, our model uncovers previously
understudied spatial and temporal heterogeneity in the so called
``democratic peace,'' whereby regime type is expected to affect the
likelihood that any two countries engage in militarized actions
against each other.  Our model also uncovers the evolving nature of
unobserved geopolitical coalitions, with memberships that conform to
theoretical expectations --- with liberal democracies aligned in one
bloc, and more authoritarian regimes aligned in another.

This paper provides applied researchers with a model that can
accommodate a variety of theorized relationships for dynamic network
outcomes that display some form of stochastic equivalence.  We make
available the open-source {\sf R} software package \if0\blind {\sf
  NetMix} \fi that implements the proposed methodology.  In the
future, we plan to further extend the model's applicability to a
variety of outcome variable types.  Similarly, and given their
prevalence in social scientific research, we plan to extend the model
to accommodate bipartite or affiliation networks.

\clearpage


\thispagestyle{empty}
\setcounter{page}{1}

\section{Online Supplementary Information for ``Dynamic Stochastic Blockmodel Regression for
  Network Data: Application to International Militarized Conflicts.''}

\clearpage
\appendix

\setcounter{page}{1}

\setcounter{table}{0}
\setcounter{equation}{0}
\setcounter{figure}{0}
\renewcommand {\theequation} {S\arabic{equation}}
\renewcommand {\thetable} {S\arabic{table}}
\renewcommand {\thefigure} {S\arabic{figure}}

\section{Marginalizing the membership vectors and the transition
probabilities}
\label{SI:collapse}

In this appendix, we show how to marginalize $\boldsymbol{\Pi}$. 
\begin{align*}
  & \int \cdots \int \prod_{t=1}^T\prod_{p\in V_{t}}
                          \left[\prod_{m=1}^M P(\boldsymbol{\pi}_{pt}\mid
    \boldsymbol{\alpha}_{ptm})^{s_{tm}}\right] \prod_{q\in
    V_{t}} P(\mathbf{z}_{p\rightarrow q,t}\mid \boldsymbol{\pi}_{pt}) P(\mathbf{w}_{p\leftarrow q,t}|\boldsymbol{\pi}_{pt}) \diff \boldsymbol{\pi}_{1}\ldots \diff\boldsymbol{\pi}_{N_t} \\
                        =  & \ \prod_{t=1}^T \prod_{p\in V_t}
                             \int\prod_{m=1}^M
                             \left[P(\boldsymbol{\pi}_{pt}\mid
                             \boldsymbol{\alpha}_{ptm})\right]^{s_{tm}}\prod_{q\in
                             V_{t}} P(\mathbf{z}_{p\rightarrow
                             q,t}\mid \boldsymbol{\pi}_{pt}) P(\mathbf{w}_{p\leftarrow q,t}|\boldsymbol{\pi}_{pt})\diff \boldsymbol{\pi}_{pt}                             \\
                         = & \ \prod_{t=1}^T \prod_{p\in V_{t}} \int
                             \prod_{m=1}^M
                             \left[\frac{\Gamma(\xi_{ptm})}{\prod_{k=1}^K \Gamma(\alpha_{ptmk})}\prod_{k=1}^K
                             \pi_{ptk}^{\alpha_{ptmk}-1}\right]^{s_{tm}}\prod_{q\in V_{t}}\prod_{k=1}^K\pi_{ptk}^{z_{p\rightarrow q,t,k}}\pi_{ptk}^{w_{p\leftarrow q,t,k}} \diff \boldsymbol{\pi}_{pt}                  \\
                         = & \ \prod_{t=1}^T \prod_{p\in V_{t}}\prod_{m=1}^M
                             \left[\frac{\Gamma(\xi_{ptm})}{\prod_{k=1}^K \Gamma(\alpha_{ptmk})}\right]^{s_{tm}}\\
                           &\quad \times \int\prod_{k=1}^K\pi_{ptk}^{\sum_{m=1}^M s_{tm}\alpha_{ptmk}-1}\prod_{q\in
                             V_{t}}\prod_{k=1}^K
                             \pi_{ptk}^{z_{p\rightarrow
                             q,t,k}}\pi_{ptk}^{w_{p \leftarrow q,t,k}}
                             \diff\boldsymbol{\pi}_{pt} 
\end{align*}
As they share a common base, we can simplify the products and define
$C_{ptk}=\sum_{q\in V_t}(z_{p\rightarrow q, t,k}+w_{p\leftarrow q,
  t,k})$ to show that the above equation is equivalent to,
\begin{equation*}
  \prod_{t=1}^T \prod_{p\in V_{t}}\prod_{m=1}^M
  \left[\frac{\Gamma(\xi_{ptm})}{\prod_{k=1}^K
      \Gamma(\alpha_{ptmk})}\right]^{s_{tm}}\int\prod_{k=1}^K\pi_{ptk}^{\sum_{m=1}^Ms_{tm}\alpha_{ptmk}
      + C_{ptk} - 1} \diff \boldsymbol{\pi}_{pt}                    
\end{equation*}
The integrand can be recognized as the kernel of a Dirichlet
distribution. As the integral is over the entire support of this
Dirichlet, we can easily compute it as the inverse of the
corresponding normalizing constant,
\begin{equation*}
        \prod_{t=1}^T \prod_{p\in V_{t}}\prod_{m=1}^M
        \left[\frac{\Gamma(\xi_{ptm})}{\prod_{k=1}^K \Gamma(\alpha_{ptmk})}\right]^{s_{tm}}\frac{\prod_{k}\Gamma(\sum_{m=1}^Ms_{tm}\alpha_{ptmk}+ C_{ptk})}{\Gamma(\sum_{m=1}^Ms_{tm}\xi_{ptm}+ 2N_{t}))}
\end{equation*}
where the sum of $C_{ptk}$ over groups $k$ is equal to twice the
number of nodes (as nodes must instantiate at least one group in each
of interactions, once as a sender and once again as a receiver) in
directed networks. A simple reorganization of factors (along with the fact that $s_{t,m}$ is an indicator vector, whereby $\sum_ms_{tm}x=\prod_mx^{s_{tm}}$) yields
equation~\eqref{eq:coll} in Section~\ref{subsec:marginal}.

\section{Details of the Collapsed Variational Algorithm}
\label{SI:VI}

\subsection{Expectation Steps}
\subsection*{E step 1: $\mathbf{Z}$ and $\mathbf{W}$}

To obtain the updates of the $\boldsymbol{\phi}_{p\rightarrow q,t}$
variational parameters, we begin by restricting equation~\eqref{eq:coll}
to the terms that depend only on $\mathbf{z}_{p\rightarrow q,t}$ (for
specific $p$ and $q$ nodes in $V_{t}$) and taking the logarithm of the
resulting expression,
\begin{align*}
  & \log P(\mathbf{Y},\mathbf{Z},
 \mathbf{W},\mathbf{S}, \mathbf{B}, \boldsymbol{\beta},
   \boldsymbol{\gamma}\mid  \mathbf{X},\mathbf{D})  \\
 = & \  z_{p\rightarrow q,t,k}\sum_{g=1}^K w_{q \leftarrow p,t,g}
     \l\{y_{pqt}\log(\theta_{pqtkh})+(1-y_{pqt})\log(1-\theta_{pqtkh})\r\}\\
  & \ + \sum_{m=1}^M s_{t m}\log\Gamma(\alpha_{ptmk}+ C_{ptg}) +\const
\end{align*}
Now, note that $C_{ptk}=C^\prime_{ptk}+z_{p\rightarrow q,t,g}$ and
that, for $x\in\{0,1\}$, $\Gamma(y + x)=y^x\Gamma(y)$. Since the
$z_{p\rightarrow q,t,k}\in \{0,1\}$, we can re-express
$\log\Gamma(\alpha_{ptmk}+ C_{ptk})=z_{p\rightarrow
  q,t,k}\log(\alpha_{ptmk}+C^\prime_{ptk}) +
\log\Gamma(\alpha_{ptmk} + C^\prime_{ptk})$ and thus simplify the expression to,
\begin{align*}
  & \ z_{p\rightarrow q,t,k}\sum_{g=1}^K w_{q\leftarrow p,t,g} \l\{y_{pqt}\log (\theta_{pqtkg})+(1-y_{pqt})\log(1-\theta_{pqtkg})\r\} \\
  & \ + z_{p\rightarrow q,t,k}\sum_{m=1}^M s_{tm}\log \left(\alpha_{ptmk}+C^\prime_{ptk}\right) + \const
\end{align*}
We proceed by taking the expectation under the variational
distribution $\widetilde{Q}$:
\begin{align*}
  &  \expec_{\widetilde{Q}}\{\log P(\mathbf{Y},\mathbf{Z},
  \mathbf{W},\mathbf{s}, \mathbf{B},
    \boldsymbol{\beta},\boldsymbol{\gamma}\mid \mathbf{D},
    \mathbf{X})\} \\
 = & \ z_{p\rightarrow q,t,g}\sum_{g=1}^K\expec_{\widetilde{Q}_2}(w_{q\leftarrow p,t,g})\bigl(y_{pqt}\log(\theta_{pqtkg})+(1-y_{pqt})\log(1-\theta_{pqtkg})\bigr) \\
 & \quad + z_{p\rightarrow q,t,g}\sum_{m=1}^M \expec_{\widetilde{Q}_1}(s_{tm})
   \expec_{\widetilde{Q}_2}\left\{\log\left(\alpha_{ptmk}+C_{ptk}^\prime\right)\right\}
   + \const
\end{align*}
The exponential of this expression corresponds to the (unnormalized) parameter vector
of a multinomial distribution
$\widetilde{Q}_2(\mathbf{z}_{p\rightarrow q,t}\mid
\boldsymbol{\phi}_{p\rightarrow q,t})$. The update for
$\mathbf{w}_{q\leftarrow p,t}$ is similarly derived.

\subsection*{E step 2: $\mathbf{S}$}

Isolating terms in Equation \ref{eq:coll} that are not constant with
respect to $s_{tm}$ for a specific $t\ne 1$ and $m$, and rolling all
other terms into a $\const$, we have
\begin{align*}
 P(\mathbf{Y},\mathbf{Z},\mathbf{W},
\mathbf{s}, \mathbf{B}, \boldsymbol{\beta},\boldsymbol{\gamma}\mid
  \mathbf{D},\mathbf{X}) 
  =  & \ \Gamma(M\eta + U_m)^{-1}\prod_{m=1}^M\prod_{n=1}^M \Gamma(\eta +
           U_{mn})
      \prod\limits_{p\in V_{t}}
       \left[\prod_{k=1}^K \frac{\Gamma(\alpha_{ptmk}+C_{ptk})}{\Gamma(\alpha_{ptmk})}\right]^{s_{tm}}\\
  & \ + \const
\end{align*}
 To isolate terms that depend on $s_{tm}$ for specific $t>1$, $m$ and $n\neq m$,
  define the following useful quantities: 
\begin{align*}
  U_m^\prime &= U_m - s_{tm}\\
  U^\prime_{mm} &= U_{mm} - s_{t-1,m}s_{tm} - s_{tm}s_{t+1,m}\\
  U^\prime_{nm} &= U_{nm} - s_{t-1,m}s_{tm}\\
  U^\prime_{mn} &= U_{mn} - s_{tm}s_{t+1,n}
\end{align*}
Focusing on the terms involving $U_m$ and $U_{mn}$, and working on a
typical case in which $1<t<T$, we can isolate parts that do not depend
on $s_{tm}$ by again recalling that, for $x\in \{0,1\}$,
$\Gamma(y+x)=y^x\Gamma(y)$:
\begin{align*}
  &\Gamma\left(M\eta + s_{tm} + U_m^\prime\right)^{-1} \Gamma(\eta + s_{t+1,m} s_{tm} + s_{t-1,m} s_{tm} + U_{mm}^\prime)\\
  & \ \times \prod^M_{n \ne m} \Gamma(\eta + s_{t+1,n} s_{tm} + U_{mn}^\prime)\Gamma(\eta + s_{tm}s_{t-1,n} + U^\prime_{nm}) \\
  =  & \ (M\eta + U_{m}^\prime)^{-s_{tm}} \Gamma(M\eta + U_{m}^\prime)^{-1} \left\{(\eta + U_{mm}^\prime +
1)^{s_{t+1,m}s_{t-1,m}}(\eta + U_{mm}^\prime)^{s_{t-1,m} - s_{t-1,m}s_{t+1,m} + s_{t+1,m}}\right\}^{s_{tm}} \\
  &\quad\times\Gamma(\eta + U_{mm}^\prime) \prod^M_{n \ne m} (\eta + U_{mn}^\prime)^{s_{t+1,n}s_{tm}} \Gamma(\eta + U_{mn}^\prime) \prod^M_{n \ne m}(\eta + U_{nm}^\prime)^{s_{tm}s_{t-1,n}}
  \Gamma(\eta + U_{nm}^\prime)
\end{align*}
at which point all $\Gamma(\cdot)$ terms are constant with respect to
$s_{tm}$ and can be rolled into the normalizing constant so that
\begin{align*}
& P(\mathbf{Y},\mathbf{Z},
\mathbf{S}, \mathbf{B}, \boldsymbol{\beta},\boldsymbol{\gamma}\mid \mathbf{D},\mathbf{X}) \\
= & \ (M\eta + U_{m}^\prime)^{-s_{tm}} \left\{(\eta + U_{mm}^\prime + 1)^{s_{t+1,m}s_{t-1,m}}(\eta + U_{mm}^\prime)^{s_{t-1,m} - s_{t-1,m}s_{t+1,m} + s_{t+1,m}}\right\}^{s_{tm}}\\ &\quad\times \prod^M_{n \ne m} (\eta + U_{mn}^\prime)^{s_{t+1,n}s_{tm}} (\eta + U_{nm}^\prime)^{s_{tm}s_{t-1,n}}\\ &\quad \times \prod_{p\in V_{t}} \left[\frac{\Gamma(\xi_{ptm})}{\Gamma(\xi_{ptm}+2N_t)}
\prod_{k=1}^K \frac{\Gamma(\alpha_{ptmk}+ C_{ptk})}{\Gamma(\alpha_{ptmk})}\right]^{s_{tm}} + \const 
\end{align*}
Taking the logarithm and expectations under the variational
distribution $\widetilde{Q}$ with respect to all variables other than
$s_{tm}$, we have,
\begin{align*}
\log \hat{\kappa}_{tm}&= -s_{tm}\expec_{\tilde{Q}_1}[\log(M\eta + U_{m}^\prime)] + s_{tm} \kappa_{t+1,m}\kappa_{t-1,m}\expec_{\tilde{Q}_1}[\log(\eta + U_{mm}^\prime + 1)]\\
                      &+ s_{tm}(\kappa_{t-1,m} -\kappa_{t-1,m}\kappa_{t+1,m} + \kappa_{t+1,m})\expec_{\tilde{Q}_1}[\log(\eta + U_{mm}^\prime)] \\
  &+ s_{tm}\sum^M_{n\neq m}\kappa_{t+1,n}\expec_{\tilde{Q}_1}[\log(\eta + U_{mn}^\prime)]\\
  &+ s_{tm}\sum^M_{n\neq m}\kappa_{t-1, n}\expec_{\tilde{Q}_1}[\log(\eta + U_{nm}^\prime)] +s_{tm} \sum\limits_{p\in V_{t}}\left[\frac{\Gamma(\xi_{ptm})}{\Gamma(\xi_{ptm}+2N_t)}\right]\\
  &+s_{tm} \sum\limits_{p\in V_{t}}\sum_{k=1}^K \expec_{\tilde{Q}}\left[\log\left[\frac{\Gamma(\alpha_{ptmk}+
    C_{ptk})}{\Gamma(\alpha_{ptmk})}\right]\right] + \const
\end{align*}
This corresponds to a multinomial distribution
$\widetilde{Q}_1(\mathbf{s}_{t}|\boldsymbol{\kappa}_{tm})$, such that
the $m$th element of its parameter vector is
\begin{align*}
 \hat{\kappa}_{tm}
  & \ \propto \ \exp\left[-\!\expec_{\widetilde{Q}_1}[\log(M\eta +  U_{m}^\prime)]\right] \exp\left[\kappa_{t+1,m}\kappa_{t-1,m}\expec_{\widetilde{Q}_1}[\log(\eta + U_{mm}^\prime + 1)]\right]\\
  &\quad \times\exp\left[(\kappa_{t-1,m} - \kappa_{t-1,m}\kappa_{t+1,m} +
    \kappa_{t+1,m})\expec_{\widetilde{Q}_1}[\log(\eta + U_{mm}^\prime)]\right]\\
  &\quad \times \prod_{n\neq m}\exp\left[\kappa_{t+1, n}\expec_{\widetilde{Q}_1}[\log(\eta + U_{mn}^\prime)]\right]\exp\left[\kappa_{t-1, n}\expec_{\widetilde{Q}_1}[\log(\eta + U_{nm}^\prime)]\right] \\
  &\quad \times \prod\limits_{p\in V_{t}}
    \left[\frac{\Gamma(\xi_{ptm})}{\Gamma(\xi_{ptm}+ 2N_t)}
    \prod_{k=1}^K \frac{\expec_{\widetilde{Q}_1}[\Gamma(\alpha_{ptmk}+
    C_{ptk})]}{\Gamma(\alpha_{ptmk})}\right]
\end{align*}
which must be normalized. When $t=T$, the
term simplifies to
\begin{align*}
    \hat{\kappa}_{Tm}
  &\propto \exp\left[-\!\expec_{\tilde{Q}_1}[\log(M\eta +
    U_{m}^\prime)]\right] \prod_{n=1}^M \exp\left[\kappa_{T-1,m}\expec_{\widetilde{Q}_1}[\log(\eta + U_{nm}^\prime)]\right]\\
  &\quad \times \prod_{p\in V_{T}}
    \left[\frac{\Gamma(\xi_{ptm})}{\Gamma(\xi_{ptm}+ 2N_t)}\prod_{k=1}^K \frac{\expec_{\widetilde{Q}_1}[\Gamma(\alpha_{pTmk}+
    C_{pTk})]}{\Gamma(\alpha_{pTmk})}\right]
\end{align*}
As before, the expectations can be approximated using a zero-order Taylor expansion.

\subsection{Maximization steps}

\subsection*{M-step 1: update for $\mathbf{B}$}

Restricting the lower bound to terms that contain $B_{gh}$, we obtain
\begin{align*}
  \mathcal{L}(\widetilde{Q})
  & \ = \ \sum^T_{t=1}\sum_{p,q\in
    E_{t}}  \sum^K_{g,h=1}\phi_{p\rightarrow q,t,g}\psi_{q\leftarrow
    p,t,h}\{y_{pqt}\log \theta_{pqtgh} +(1-y_{pqt})\log(1-\theta_{pqtgh})\}\\
&\qquad- \sum^K_{g,h=1}\frac{(B_{gh}-\mu_{gh})^2}{2\sigma_{gh}^2} + \const
\end{align*}
We optimize this lower bound with respect to $\mathbf{B}_{gh}$ using a
gradient-based numerical optimization method. The corresponding
gradient is given by,
\begin{align*}
  \frac{\partial\mathcal{L}_{B_{gh}}}{\partial B_{gh}}
  & \ = \ \sum^T_{t=1} \sum _{p,q\in E_{t}}  \phi_{p\rightarrow q,t,g} \psi_{q\leftarrow p,t,h}\left(y_{pqt}-\theta_{pqtgh}\right)- \frac{B_{gh}-\mu_{B_{gh}}}{\sigma_{B_{gh}}^2}
\end{align*}

\subsection*{M-step 2: update for $\boldsymbol{\gamma}$}

Restricting the lower bound to terms that contain
$\boldsymbol{\gamma}$, and recalling that
$\theta_{pqtgh}=[1+\exp(-B_{gh}-\mathbf{d}_{pqt}\boldsymbol{\gamma})]^{-1}$, 
we have
\begin{align*}
  \mathcal{L}(\tilde{Q})
  & \ = \ \sum\limits^T_{t=1}\sum_{p,q\in E_{t}} 
    \sum^K_{g,h=1} \phi_{p\rightarrow q,t,g}\psi_{q\leftarrow
    p,t,h}\l\{y_{pqt}\log \theta_{pqtgh}+(1-y_{pqt})\log(1-\theta_{pqtgh})\r\}\\
&\quad -\sum^{J_d}_{j}\frac{(\gamma_{j}-\mu_\gamma)^2}{2\sigma^2_\gamma} + \const
\end{align*}
To optimize this expression with respect to $\gamma_{j}$ (the $j$th
element of the $\boldsymbol{\gamma}$ vector), we again use a numerical
optimization algorithm based on the following gradient,
\begin{align*}
  \frac{\partial \mathcal{L}(\widetilde{Q})}{\gamma_{j}}
  & \ = \  \sum^T_{t=1} \sum_{p,q\in E_{t}}  \sum^K_{g,h=1}
      \phi_{p\rightarrow q,t,g} \psi_{q\leftarrow p,t,h} d_{pqtj}\left(y_{pqt}-\theta_{pqtgh}\right)- \frac{\gamma_{j}-\mu_{\gamma}}{\sigma_{\gamma}^2}
\end{align*}

\subsection*{M-step 3: update for $\boldsymbol{\beta}_m$}

Let $\alpha_{ptmk}=\exp\l(\mathbf{x}_{pt}^\top\boldsymbol{\beta}_{km}\r)$ and $\xi_{ptm}=\sum_{k=1}^K\alpha_{ptmk}$. To find the optimal
value of $\boldsymbol{\beta}_{km}$, we roll all terms not involving the coefficient vector into a constant:
\begin{align*}
  \mathcal{L}(\widetilde{Q})
  & \ = \ \sum^T_{t=1}\sum^M_{m=1}\kappa_{tm}\sum_{p\in V_{t}}\l[\log\Gamma(\xi_{ptm})-\log\Gamma(\xi_{ptm}+2N_t)\r] \\
  & \ \quad + \sum^T_{t=1}\sum^M_{m=1}\kappa_{tm}\sum_{p\in V_{t}}\sum^K_{k=1}\left[\expec_{\widetilde{Q}_2}[\log\Gamma(\alpha_{ptmk}+C_{ptk})]-\log\Gamma(\alpha_{ptmk})\right]\\
 &\qquad - \sum_{k=1}^K\sum^M_{m=1}\sum^{J_x}_{j=1} \frac{(\beta_{mkj}-\mu_\beta)^2}{2\sigma^2_\beta} + \const
\end{align*}
 No closed
form solution exists for an optimum w.r.t. $\beta_{mkj}$, but a gradient-based
algorithm can be implemented to maximize the above expression. The corresponding
gradient with respect to each element of $\boldsymbol{\beta}_{mk}$ is
given by,
\begin{align*}
  \frac{\partial\mathcal{L}(\tilde{Q})}{\partial \beta_{mkj}}
  & \ = \ \sum^T_{t=1}\kappa_{tm}\sum_{p\in V_{t}}\alpha_{ptmk}x_{ptj}\Bigl(\expec_{\widetilde{Q}_2}[\breve{\psi}(\alpha_{ptmk}+C_{ptk})-\breve{\psi}(\alpha_{ptmk})]\\
  & \  \qquad + \l[\breve{\psi}(\xi_{ptm})-\breve{\psi}(\xi_{ptm}+2N_t)\r]\Bigr)\\
  & \ \qquad - \frac{\beta_{mkj} - \mu_\beta}{\sigma_\beta^2}
\end{align*}
where $\breve{\psi}(\cdot)$ is the digamma function. Once again,
we can approximate expectations of non-linear functions of random
variables using a zeroth-order Taylor series expansion. As is the case
of the multinomial logit model, we set $\boldsymbol{\beta}_{1,m}\equiv 0\; \forall m$, making group 1 a reference for identification
purposes.

\section{A Simulation Study}
\label{sec:simulation}

Using synthetic dynamic networks, we evaluate the estimation accuracy
with respect to the mixed-membership vectors and the blockmodel
matrices under three scenarios: easy, medium, and hard learning
problems. We also examine the quality of regression coefficient
estimates, and the ability of the model to recover the parameters
associated with the underlying HMM. Finally, we compare the results of
fitting a fully specified \dynMMSBM{} and fitting a separate MMSBM
(without covariates) to each time period, showing the substantial
gains in error reduction resulting from the use of our proposed model.

Our synthetic networks are composed of 100 nodes observed
over $t\in \{1, \ldots, 9\}$ time periods, and are constructed as follows:
\begin{enumerate}
\item  For each node $pt$ and dyad $pqt$ at time $t>1$, generate a single monadic and dyadic predictor using a random walk, so that
  ${x_{pt}= x_{p,t-1} +\epsilon_{xt}}$,
  ${d_{pqt}=d_{pq,t-1}+\epsilon_{dt}}$, with 
  ${x_{p1}\sim N\left(0, 2\right)}$,
  ${d_{pq,1}\sim N\left(0, 2\right)}$, and
  $\epsilon_{xt}\sim N(0,1)$, $\epsilon_{dt}\sim N(0,1)$.
\item For each node at time $t$, sample a 2-dimensional
  mixed-membership vector from a 2-component mixture of Dirichlet
  distributions, so that
  \[
    \boldsymbol{\pi}_{pt}\sim\prod\limits_{m=1}^2\left[\text{Dirichlet}\left(\exp\left(\mathbf{x}_{pt}^\top\boldsymbol{\beta}_{m}\right)\right)\right]^{s_{tm}}
\]
where $\mathbf{x}_{pt} = [1\; x_{pt}]^\top$, and $s_{tm}$ indicates a state
$m\in \{1,2\}$ of the hidden Markov process, such that $s_{t1}=1$ for
$t\in \{1, \ldots, 5\}$, $s_{t2}=1$ for $t\in \{6,\ldots, 9\}$, and
$s_{tm}=0$ otherwise (i.e. there is a changepoint in the underlying left-to-right
HMM between time-points 5 and 6).
\item For each node $pt$ and $qt$ in directed dyad $pqt$, sample a pair of
  group memberships
  \[
    z_{pt\rightarrow qt}\sim
    \text{Categorical}\left(\boldsymbol{\pi}_{pt}\right) \text{ and }
    w_{qt\leftarrow pt}\sim\text{Categorical}(\boldsymbol{\pi}_{qt})
    \]
\item Finally, and for the same dyad, sample an edge
  \[
    y_{pqt}\sim\text{Bernoulli}\left(\text{logit}^{-1}\left(B_{z_{pt\rightarrow qt}, w_{qt\leftarrow pt}} + \gamma_1d_{pqt}\right)\right)
  \]
  where $\gamma_1=0.1$. 
\end{enumerate}

To explore the conditions under which the model performs best, as well
as those under which learning the model's various parameters
can be particularly challenging, we refine this data-generating
process by defining three sets of values for $\bm{B}$ and $\bm{\beta}$
designed to generate easy, medium, and hard learning
scenarios. They differ in the extent to which memberships are truly mixed
(with more clearly defined memberships being easier to
learn), and with respect to the extent to which the blockmodels
generate distinct equivalence classes of nodes (with more clearly
defined block structures being easier to learn). Accordingly,
each scenario's DGP is completed using the parameters in presented in
Table~\ref{tab:params}.

\begin{table}[hb!t]  
  \begin{tabular}{r c  c  c }
\toprule
& \textbf{Easy} & \textbf{Medium} & \textbf{Hard}\\ 
\midrule
$g^{-1}(\mathbf{B})=$ & $\begin{bmatrix} 0.85 & 0.01\\0.01 & 0.99 \end{bmatrix}$ & $\begin{bmatrix}0.65 & 0.35\\0.20 & 0.75\end{bmatrix}$ &$\begin{bmatrix}0.65 & 0.40\\0.50 & 0.45\end{bmatrix} $\\[0.75cm]
$\boldsymbol{\beta}_1=$&$\begin{bmatrix*}[r]-4.5&-4.5\\0.0&0.0\end{bmatrix*}$&$\begin{bmatrix*}[r]0.05&0.75\\-0.75&-1.0\end{bmatrix*}$&$\begin{bmatrix*}[r]0.0&0.0\\-0.75&-1.0\\\end{bmatrix*}$\\[0.75cm]
$\boldsymbol{\beta}_2=$&$\begin{bmatrix*}[r]-4.5&-4.5\\0.0&0.0\end{bmatrix*}$&$\begin{bmatrix*}[r]-0.05&0.55\\-0.75&0.75\end{bmatrix*}$&$\begin{bmatrix*}0.0&0.0\\-0.75&0.75\\\end{bmatrix*}$\\
\bottomrule
\end{tabular}
\caption{\textbf{Parameters in three different dynamic network
    DGPs}. The three columns correspond to three types of networks,
  varying in terms of inferential complexity. In turn, the rows
  contain the corresponding values of the blockmodel $\mathbf{B}$ and
  the regression coefficient vectors $\bm{\beta}$, one for each state
  of the HMM.}
\label{tab:params}
\end{table}

\begin{figure}[ht!b] \spacingset{1}
  \centering
  \includegraphics[scale=0.65]{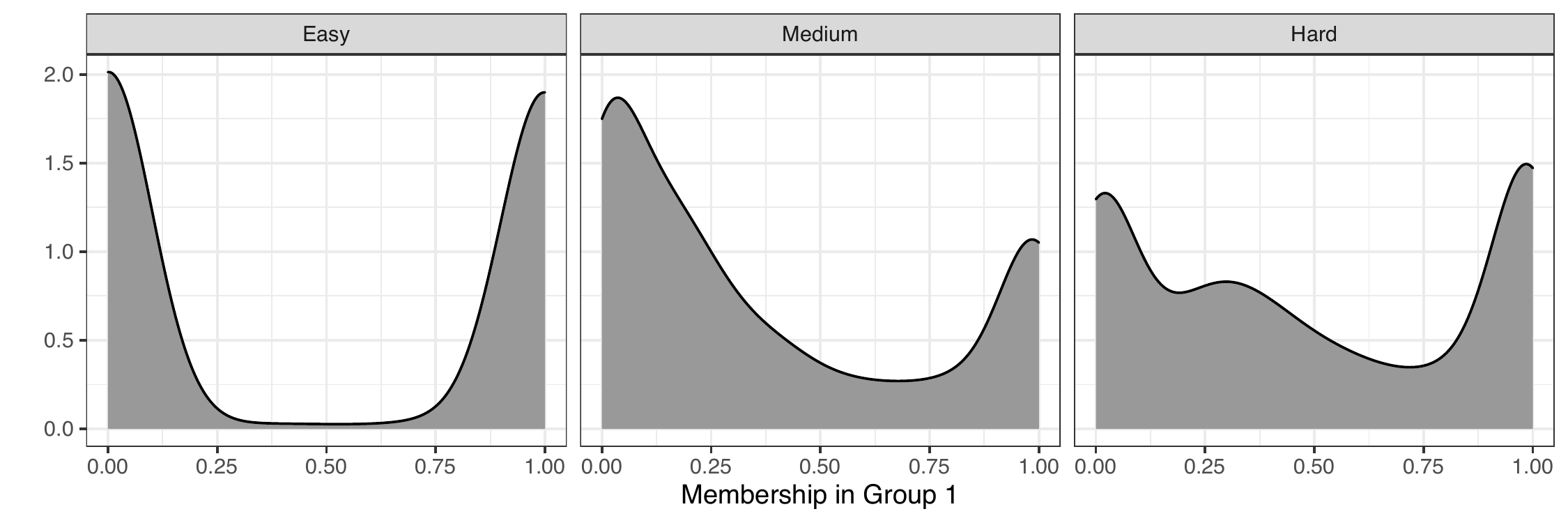}
  \caption{\textbf{Simulated mixed-memberships in synthetic
      networks}. The plots depict the mixed-membership vectors of
    nodes in three simulated networks, each with 100 nodes observed
    over 9 time periods. It shows the memberships of nodes in networks generated under an
    `easy' DGP (i.e. one where memberships are not mixed, and in which
  the block structure is clear), `hard' (i.e. one where memberships
  are extremely mixed, and no block structure is apparent in the
  network) and `medium' (i.e. where some nodes display a mixture of
  group memberships, and a block structure is somewhat apparent in the
  network) on the left, right, and central panels, respectively.}
  \label{fig:ternary}
\end{figure}

Generating a single, 9-period network under each of these scenarios results in the
mixed memberships depicted in Figure~\ref{fig:ternary}, which shows
the density of membership into the first of two groups across all
nodes and time periods. While the `easy' scenario has very clearly
defined memberships of most nodes into one of the underlying groups,
the `hard' scenario has a substantial number of nodes whose membership
is decidedly more mixed. The medium, more `realistic' scenario has a
non-negligible number of nodes whose membership is mixed, and a
distinct group imbalance in favor of the second group.  

\subsection{Accuracy of estimation: mixed-memberships and blockmodels}

\begin{figure}[t!] \spacingset{1}
  \centering
  \includegraphics[scale=0.65]{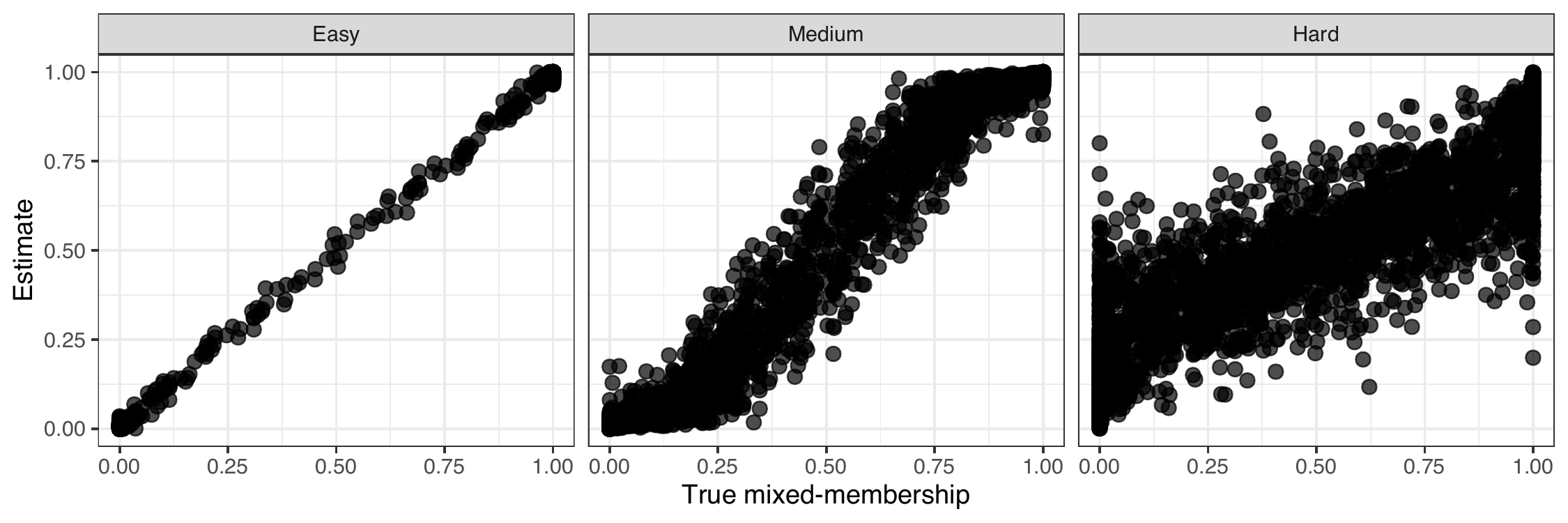}
  \includegraphics[scale=0.65]{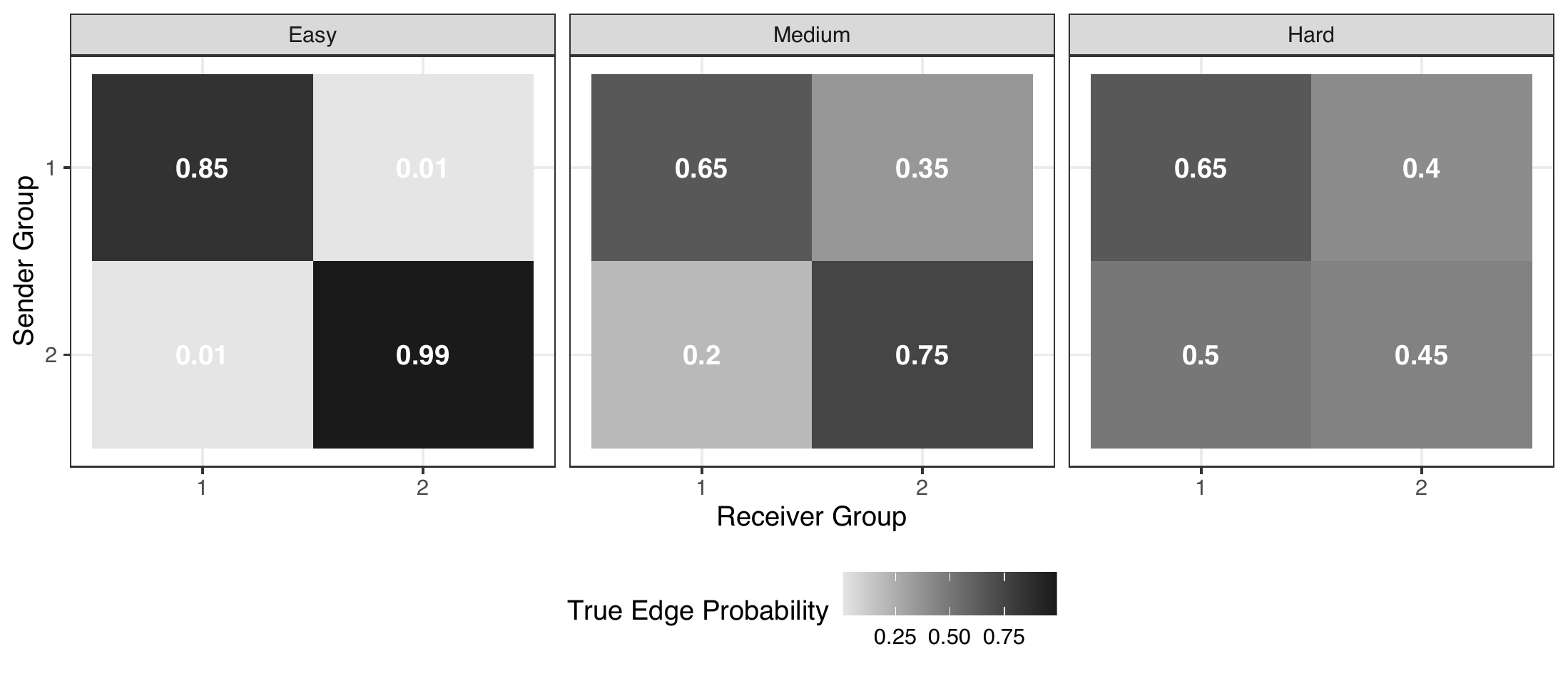}
  \caption{\textbf{Estimation accuracy}. For each DGP scenario, the figure
    shows the estimated mixed-membership vectors (top row) and the estimated blockmodels (bottom row)
    against their known values (indicated by the white numbers in each cell of the blockmodel
    for the bottom row). Overall,
    accuracy of retrieval both sets of parameters depends on the
    complexity of the learning problem, although recovery is generally
  very good, even under `hard' inferential conditions.}
  \label{fig:truevest}
\end{figure}

Overall, and as expected, the accuracy with which \dynMMSBM{} can
retrieve the true mixed-membership vectors depends on the problem's
complexity. The top panel of Figure~\ref{fig:truevest} shows the estimated mixed-membership
values against their known, true values, evidencing a decrease in
estimation accuracy as we move from an easy to a hard inferential
task. Despite the clear deterioration, \dynMMSBM{} is still able to
produce good quality estimates even under hard inferential situations,
with estimates that have a 0.94 correlation with their true
values.

The model is also able to accurately estimate the blockmodel
structure, as the bottom row of Figure~\ref{fig:truevest} reveals. For
each cell of the blockmodels, the true probability of an edge between
members of any two groups is shown in white letters, while the cell
itself is colored in accordance to the corresponding estimated
values. Once again, and although the quality of these estimates (predictably)
decreases as the inferential complexity of the scenario increases, the estimation
error remains low.    

\subsection{Estimation accuracy: regression coefficients}

The two most distinctive features of the proposed model are its
ability to incorporate predictors of the mixed-membership vectors and
to account for network dynamics. We evaluate the accuracy with which
our proposed estimation strategy recovers known parameter values. To
do so, we simulate 100 replicates of the 9-period network described
above, generated under our medium, more `realistic' DGP and holding all design
matrices constant across replicates. After generating all 50 networks, we
use our model to obtain estimates of the effect of the monadic
predictor on block memberships, as well as of the marginal probability
that the hidden Markov process is in either of the two states for each
time period.

\begin{figure}[t!] \spacingset{1}
  \centering
  \includegraphics[scale=0.65]{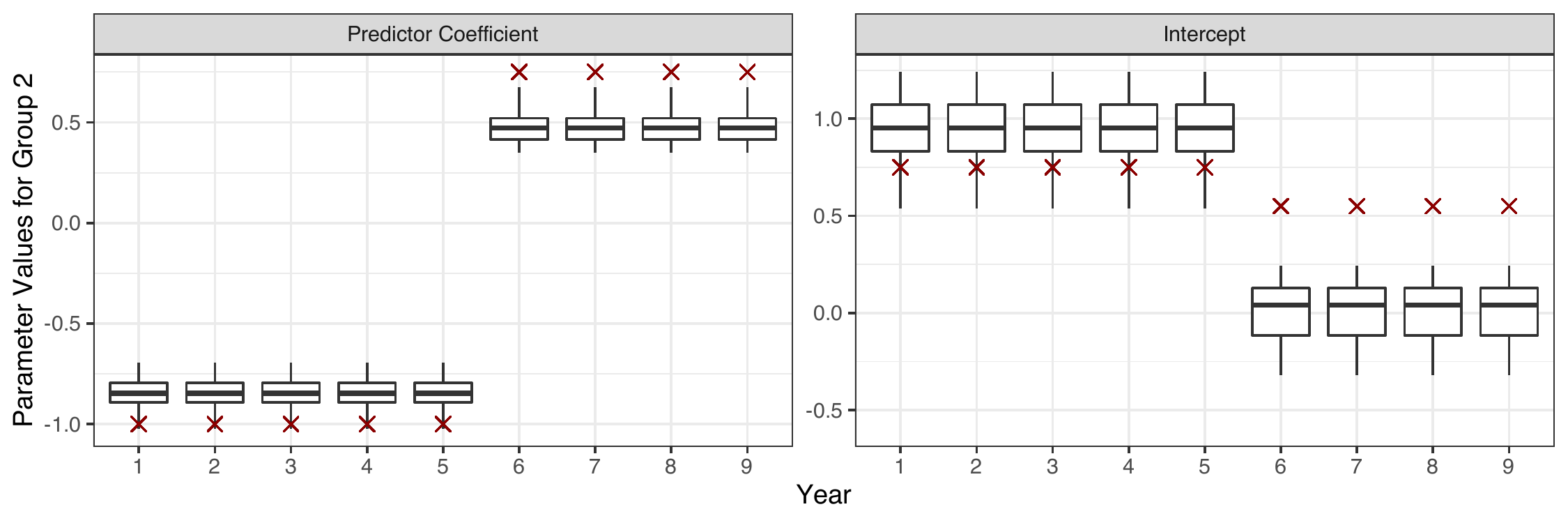}
  \caption{\textbf{Estimated parameters of block membership regression}. The figure shows, for each time period, the
    HMM-weighted effect of a continuous predictor on the probability
    of instantiating latent group 2 (left panel), and the HMM-weighted intercept of the corresponding regression line (right panel),
    estimated on 100 networks generated according to our medium DGP. In each
    instance, the red ``x'' indicates the true parameter value for that
    time period, given a known HMM state.}
  \label{fig:SynthFX}
\end{figure}

Figure \ref{fig:SynthFX} shows, for each time period, the distribution
of estimated effect sizes of the monadic predictor and intercepts for
the regression of membership into the second latent group (as
boxplots), along with the true parameter values (shown as a red ``x''). We
obtain estimates for each time period by computing the weighted
average of estimated parameters in the two hidden Markov states, using
estimated marginal probabilities over states in each time period as
our weights. The model is typically able to identify the underlying
Markov state that generated the networks, which in turn translates
into correctly estimated (albeit regularized) effects of the monadic
covariate on membership probabilities. Quality of recovery for
regression parameters associated with a given block depends
heavily on the extent to which that block is commonly instantiated in
the network. And although changes in intercepts across time periods
are also correctly recovered, the intercepts themselves tend to be
overestimated. This phenomenon, which we found to be common in all our
simulations, is likely the result of the difficulty in pinning down
the precision of the latent membership vectors. Despite these issues,
the mean of the memberships is correctly recovered (as shown earlier in
Figure~\ref{fig:truevest}).  

\subsection{Comparison to non-dynamic MMSBM}

\begin{figure}[t!] \spacingset{1}
  \centering
  \includegraphics[scale=0.75]{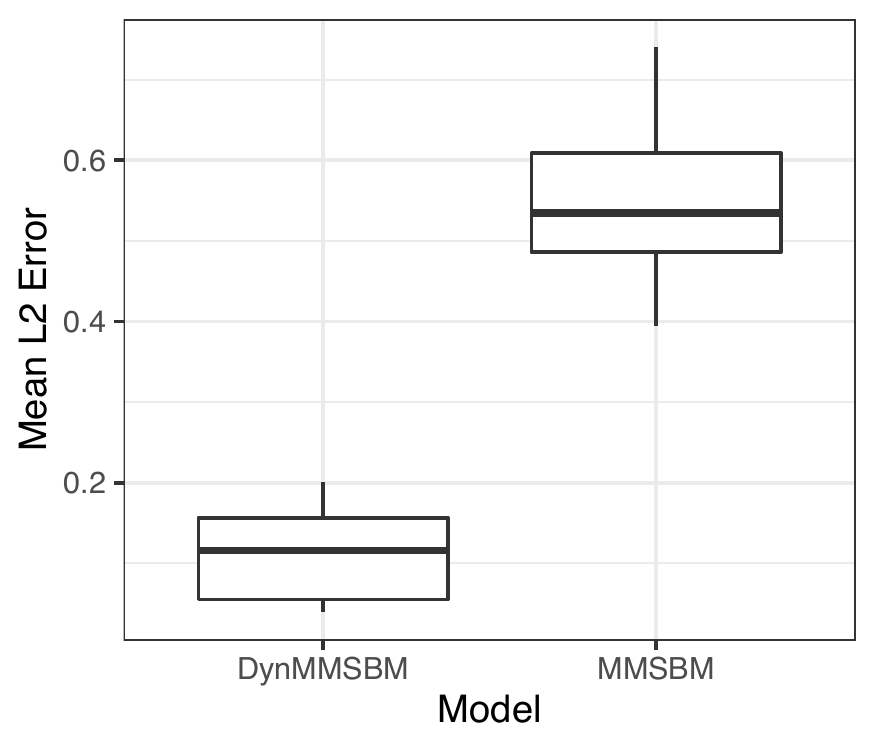}
  \caption{\textbf{Error for estimated mixed membership vectors}. The figure
    shows average $L_2$ distances between estimated and true
    mixed-membership vectors for all nodes in each of 100 replicated
    dynamic networks. On the left, estimates are generated using
    \dynMMSBM{}. On the right, estimates are generated using the
    canonical MMSBM, fit separately to the nine time periods in each
    simulated network.}
  \label{fig:SynthError}
\end{figure}

Finally, and to further evaluate the benefits of modeling the dynamic
nature of the network, we estimate a separate MMSBM model to the
networks in each time period, and compare their estimated
mixed-memberships to those of a single \dynMMSBM{} estimated on the full
set of networks. In both cases, we omit all
covariates, but estimate the $\bm{\alpha}_{ptm}$ parameters associated
with the mixed-membership vectors. After estimating both sets of
models on each of the 100 replications of the ``medium'' networks,
we compute the average $L_2$ error in estimated mixed-memberships
across nodes. The results are presented in Figure~\ref{fig:SynthError}.

In general, \dynMMSBM{} performs consistently better than the MMSBM
estimated on each time period, and the latter shows much more
variability in terms of accuracy. A major challenge for the per-year
approach consists of realigning the estimated group labels, which
(under the assumptions of our model) should be done by realigning the
cells of the blockmodel, as all other parameters (such as the
mixed-memberships themselves) are subject to change overtime. Being
estimated using just a fraction of the data, however, the blockmodels
obtained in the per-year approach prove too noisy to be useful in the
realignment exercise, thus contributing to the variable
accuracy of the non-dynamic approach. In contrast, \dynMMSBM{} is able
to recover the underlying blockmodel much more accurately, thus
contributing to the correct estimation of the latent memberships
across simulations.

\section{Additional Empirical Results}
\label{app:additional}

\subsection{Model forcast accuracy results, with different numbers of latent groups}

Table~\ref{tb:groups_AUC} presents out-of-sample (forcast) errors for
models with different numbers of latent groups. We estimate models for
networks observed between 1816 and 2010, and compute the area under
the receiver operating characteristic curve for forecast networks in
2009 and 2010. In contrast, the results reported in the main text all
use the entire time span (1816-2010).
\begin{table}[h!t]
\spacingset{1}
\centering
\begin{tabular}{ll}
  \hline
\# Groups & AUROC  \\ 
  \hline
2 & 0.966  \\ 
 & (0.020)  \\ 
  3 & 0.989  \\ 
   & (0.012)  \\ 
  4 & 0.986  \\ 
   & (0.013)  \\ 
  5 & 0.984  \\ 
   & (0.014)  \\ 
  6 & 0.986  \\ 
   & (0.013)  \\ 
  7 & 0.975  \\ 
   & (0.017)  \\ 
   \hline
\end{tabular}   \spacingset{1}
\caption{\textbf{Out of Sample Prediction, Different Latent Groups}. The table displays the area under the ROC curve (AUROC) and associated standard error for specifications with 2-7 Latent Groups.  Each model is fit on data from 1816-2008 and used to forecast conflict in the period 2009-2010.} \label{tb:groups_AUC}
\end{table}

\clearpage
\subsection{Model including geographic region indicators}

\begin{figure}[h!t] \spacingset{1}
  \centering
  \includegraphics[scale=.58]{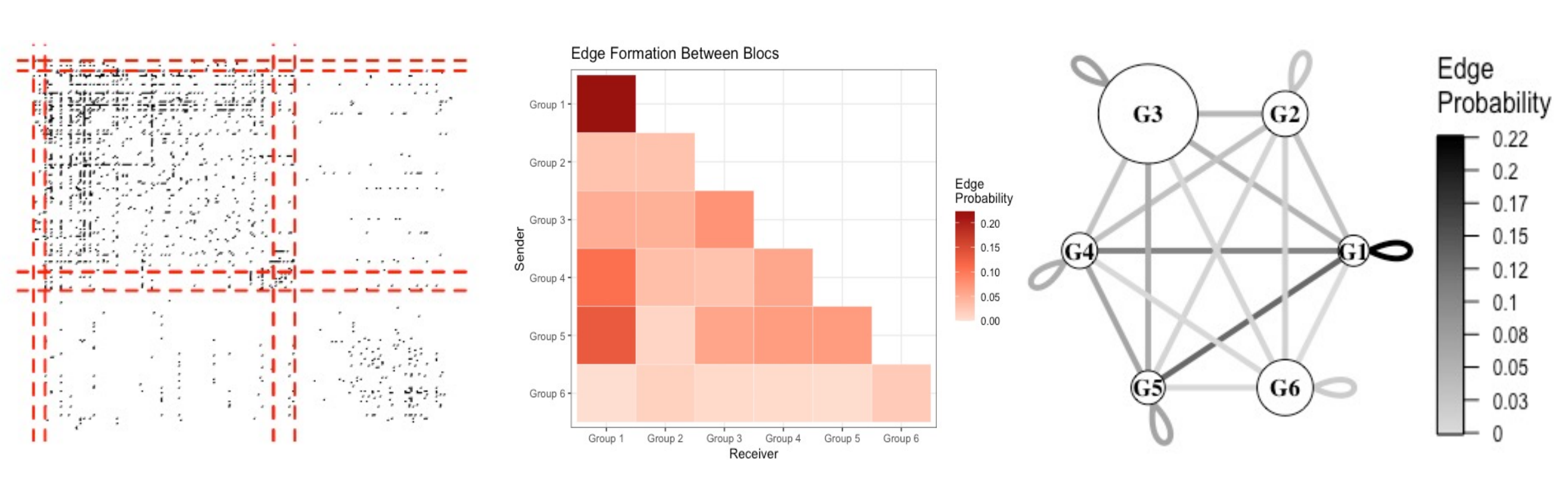}
  \caption{\textbf{Estimated blockmodel in the conflict network
      (regional model)}. Blockmodel visualizations for a specification
    including an indicator for state region.  The left panel displays
    the adjacency matrix of militarized disputes.  The middle panel
    displays the estimated probability of conflict between groups as a
    heat map.  The right panel is a network graph summarizing the
    estimated blockmodel.  The blockmodel in this specification is
    moderately correlated (0.104) with the primary model.}
  \label{fig:blockmodel_regions}
\end{figure}

\begin{figure}[!t] \spacingset{1}
     \includegraphics[scale=.65]{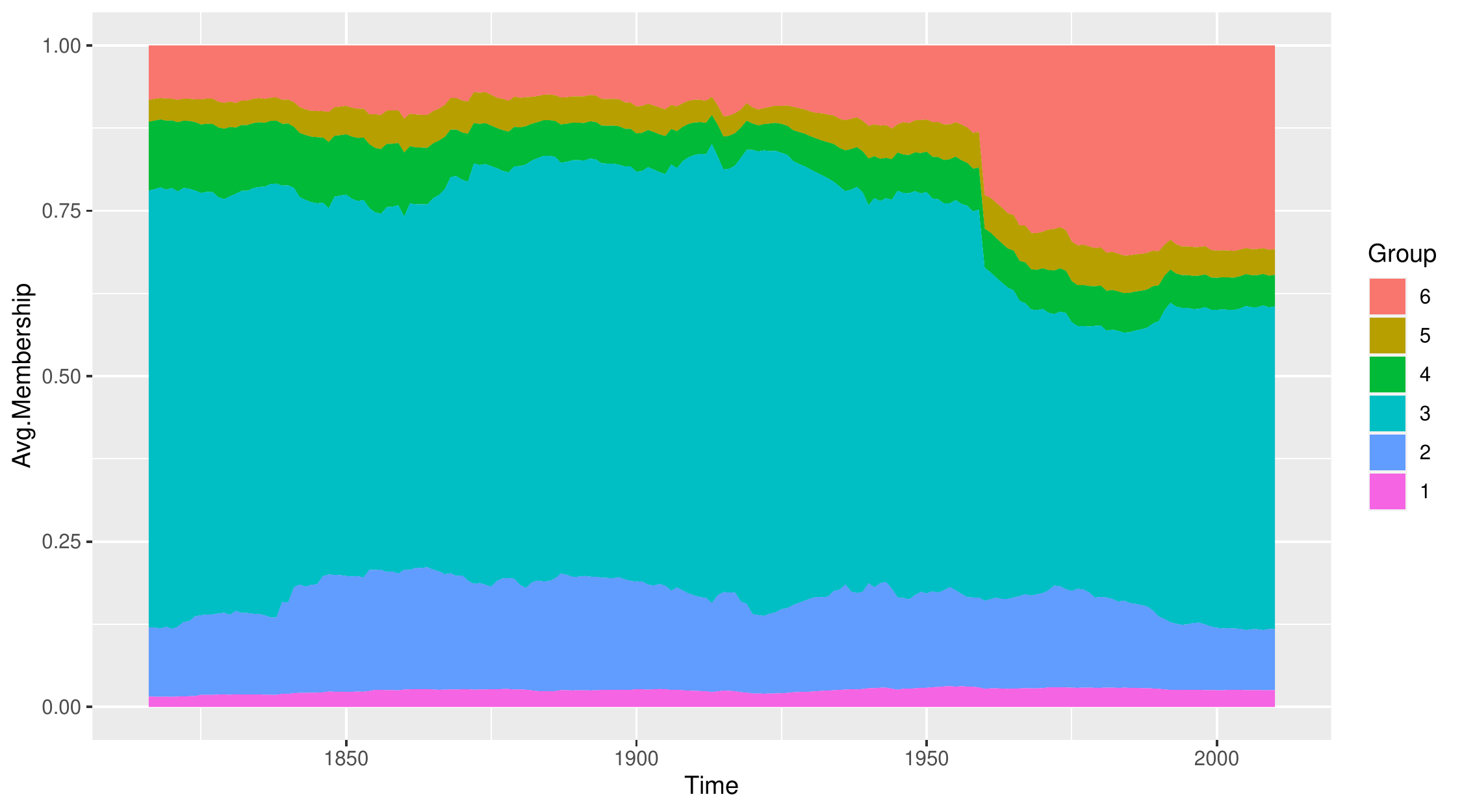} 
\vspace{-.2in}
\caption[cluster time]{\textbf{Membership in Latent Groups over Time
    (regional model)}. The figure shows the average proportion of
  membership in six latent groups for each year from 1816--2010.  The
  estimated evolution of membership in this specification is similar
  in some respects to the primary specification (e.g., the sizeable increase in Group~6 membership in later decades), but differs in
  others (e.g., Group~2 membership is smaller in earlier years).  Notably,
  this specification does not experience transitions in the hidden
  Markov state.  This may be attributable to the addition of
  indicators for state region, which are static over time.}
  \label{fig:cluster_time_regions}
\end{figure}

As an alternative way of capturing geographic determinants of
militarized disputes, we estimate a model that includes regional
indicators as predictors.Figure~\ref{fig:blockmodel_regions} presents
the estimated blockmodel when using this alternative specification. In
turn, Figure~\ref{fig:cluster_time_regions} shows the evolution of
memberships into estimated latent groups under this
specification. Finally, Table~\ref{tb:CWstates_region} shows countries
with the highest average estimated membership in each of the groups during the Cold War for this alternative specification. Many group assignments are
similar to the main specification.  Nodes with high membership in
Groups 4-6, for example, reflect many of the same countries as in the
specification reported in the main text. The most notable difference
in this specification is the grouping of Western and some Eastern bloc
countries into a single, more belligerent latent group (Group~3):
the United States, United Kingdom, and West Germany are part of this group along with Russia and Poland (not shown).

\vspace{14mm}

\begin{table}[h!t]
  \spacingset{1.1}
  \centering
  \begin{tabular}{lll}
    \hline
    \textbf{Group 1}     & \textbf{Group 2}  & \textbf{Group 3}       \\
    \hline
0.083 Lebanon & 0.571 Paraguay & 0.998 USA \\ 
0.079 North Yemen & 0.46 Bolivia & 0.996 India \\ 
0.075 Yemen & 0.377 Argentina & 0.992 Canada \\ 
0.074 South Yemen & 0.357 Ecuador & 0.976 West Germany \\ 
0.074 Bhutan & 0.348 Uruguay & 0.975 UK \\ 
0.073 Tunisia & 0.345 Chile & 0.974 Germany \\ 
0.073 Libyan & 0.332 Peru & 0.966 Italy \\ 
0.073 Jordan & 0.326 Mongolia & 0.963 Japan \\ 
0.072 Maldives & 0.324 Korea North & 0.959 France \\ 
0.072 Kuwait & 0.324 Albania & 0.955 Pakistan \\ 
0.072 United Arab Emirates & 0.303 Cambodia & 0.947 Sri Lanka \\ 
0.07 Morocco & 0.289 Taiwan & 0.944 Turkey \\ 
0.07 Syria & 0.267 Laos & 0.94 Russia \\ 
0.069 Algeria & 0.252 Vietnam & 0.937 Netherlands \\ 
0.069 Iraq & 0.246 Myanmar & 0.937 Belgium \\ 
    \hline
    \textbf{Group 4}     & \textbf{Group 5}  & \textbf{Group 6}       \\
    \hline
0.212 Bhutan & 0.282 Bahrain & 0.948 Seychelles \\ 
0.158 Maldives & 0.282 Oman & 0.941 Sao Tome Principe \\ 
0.155 Bahrain & 0.281 Qatar & 0.927 Zanzibar \\ 
0.154 Oman & 0.269 Jordan & 0.886 Liechtenstein \\ 
0.154 Tunisia & 0.268 Saudi Arabia & 0.878 Comoros \\ 
0.153 Qatar & 0.266 United Arab Emirates & 0.875 Gambia \\ 
0.153 Jordan & 0.264 Tunisia & 0.863 Cape Verde \\ 
0.153 Kuwait & 0.264 Kuwait & 0.842 Equatorial Guinea \\ 
0.153 South Yemen & 0.263 South Yemen & 0.841 Swaziland \\ 
0.152 United Arab Emirates & 0.259 Libyan & 0.834 Botswana \\ 
0.15 Libyan & 0.258 Algeria & 0.823 St Kitts-Nevis \\ 
0.149 North Yemen & 0.249 Morocco & 0.82 Lesotho \\ 
0.149 Saudi Arabia & 0.247 Djibouti & 0.798 Gabon \\ 
0.147 Algeria & 0.246 North Yemen & 0.797 Dominica \\ 
0.145 Morocco & 0.236 Syria & 0.796 Antigua-Barbuda \\ 
    \hline
  \end{tabular}
  \caption{\textbf{States with Highest Membership in Latent Groups,
      Cold War period (regional model)}.  Average group membership in
    the years 1950-1990 is reported beside the state name for the top
    15 states in each latent group.}
 \label{tb:CWstates_region}
\end{table}

\clearpage
\subsection{Models estimated on entire dataset, with different  numbers of latent groups}

\begin{figure}[h!t] \spacingset{1}
  \centering
  \includegraphics[scale=.54]{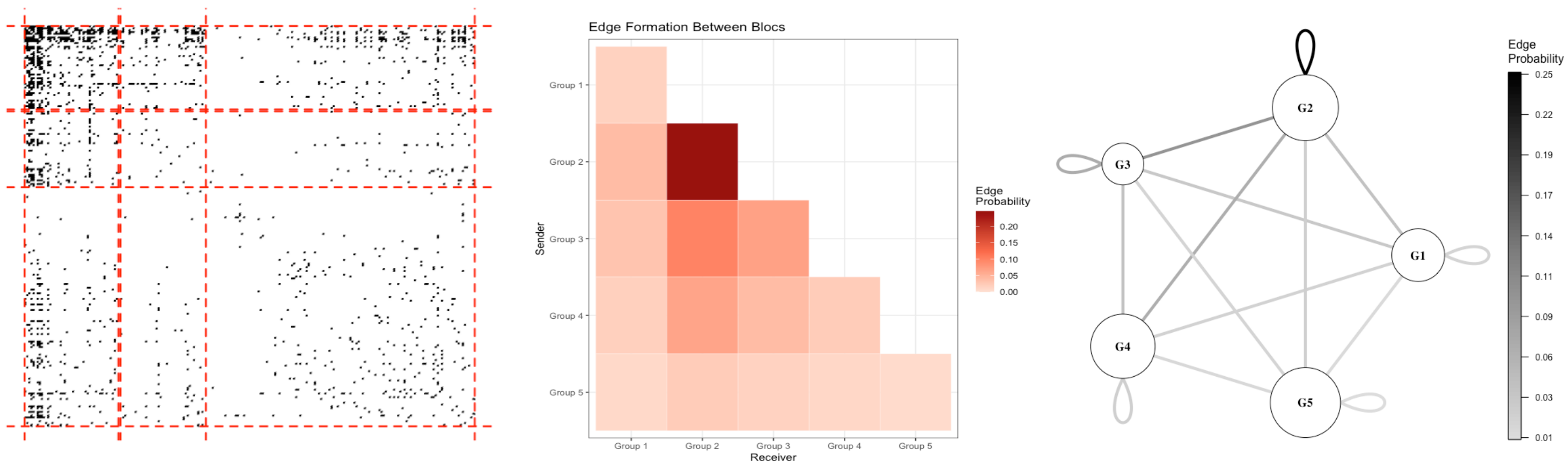}
  \vspace{-0.2in}
  \caption{\textbf{Estimated blockmodel in the conflict network
      (5-group specification)}. The left panel displays the adjacency
    matrix of militarized disputes between 216 states.  The middle
    panel displays the estimated probability of conflict between
    groups as a heat map.  The right panel is a network graph
    summarizing the estimated blockmodel.}
  \label{fig:blockmodel5}
\end{figure}

\begin{figure}[h!t] \spacingset{1}
  \centering
  \includegraphics[scale=.54]{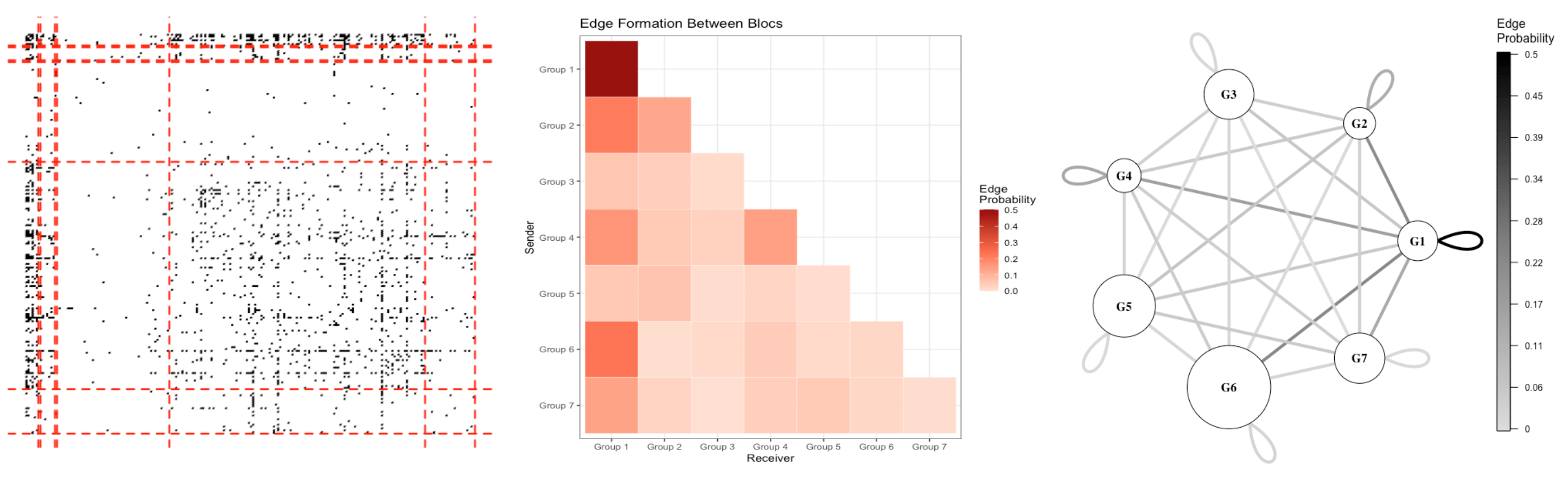}
  \vspace{-0.2in}
  \caption{\textbf{Estimated blockmodel in the conflict network
      (7-group specification)}. The left panel displays the adjacency
    matrix of militarized disputes between 216 states.  Dotted lines
    separate states by estimated group membership; some groups are not
    visible in the adjacency matrix since they have very low
    membership.  The middle panel displays the estimated probability
    of conflict between groups as a heat map.  The right panel is a
    network graph summarizing the estimated blockmodel.}
  \label{fig:blockmodel7}
\end{figure}

Figures~\ref{fig:blockmodel5} and~\ref{fig:blockmodel7} show the
estimated blockmodels of models defined with 5 and 7 latent groups,
respectively. The blockmodels mostly reflect similar relationships and
memberships to those of the model reported in the main text, which
uses 6 latent groups.

\clearpage
\subsection{Estimated blockmodel, in table format}

Table~\ref{tb:blockmodel} presents the estimated blockmodel displayed
in graphical form in Figure 1 of the main text.

\begin{table}[h!t]
\centering
\begin{tabular}{rrrrrrr}
  \hline
 & Group 1 & Group 2 & Group 3 & Group 4 & Group 5 & Group 6 \\ 
  \hline
Group 1 & $0.182$ & $0.137$ & $0.017$ & $0.004$ & $0.180$ & $0.047$ \\ 
Group 2 & $0.137$ & $0.105$ & $0.068$ & $0.028$ & $0.018$ & $0.011$ \\ 
Group 3 & $0.017$ & $0.068$ & $0.014$ & $0.017$ & $0.011$ & $0.003$ \\ 
Group 4 & $0.004$ & $0.028$ & $0.017$ & $0.009$ & $0.001$ & $0.004$ \\ 
Group 5 & $0.180$ & $0.018$ & $0.011$ & $0.001$ & $0.049$ & $0.044$ \\ 
Group 6 & $0.047$ & $0.011$ & $0.003$ & $0.004$ & $0.044$ & $0.030$ \\ 
   \hline
\end{tabular}   \spacingset{1}
\caption{\textbf{Group-Level Edge Formation Probabilities}.  The table
  displays the probability of interstate conflict between nodes that
  instantiate membership in each of six latent groups.  The diagonal
  shows rates of intra-group conflict and off-diagonal shows rates of
  conflict between groups.} \label{tb:blockmodel}
\end{table}

\clearpage
\subsection{Countries with highest membership probabilities}
Table~\ref{tb:CWstates} shows, for each of the 6 estimated latent
groups in the main text, the countries with the highest average
membership probabilities during the 1950-1990 time period. The group
assignments are consistent with known geopolitical coalitions in the
Cold War, with Western allies in Group 1, Eastern bloc countries
clustered in Group 2, Western-leaning neutral states in Group 3, and
states engulfed in proxy conflicts in Group 4.

\begin{table}[h!t]
  \spacingset{1.1}
  \centering
  \begin{tabular}{lll}
    \hline
    \textbf{Group 1}     & \textbf{Group 2}  & \textbf{Group 3}       \\
    \hline
0.158 USA & 0.972 Russia & 0.888 Costa Rica \\ 
0.138 UK & 0.969 China & 0.88 New Zealand \\ 
0.138 Japan & 0.827 Germany East & 0.878 Ireland \\ 
0.137 India & 0.81 Poland & 0.874 Jamaica \\ 
0.132 West Germany & 0.766 Czechoslovakia & 0.868 Norway \\ 
0.115 Italy & 0.763 Korea North & 0.866 Finland \\ 
0.109 France & 0.747 Romania & 0.863 Denmark \\ 
0.100 Canada & 0.744 Iran & 0.862 Switzerland \\ 
0.082 Belgium & 0.743 Indonesia & 0.859 Luxembourg \\ 
0.081 Australia & 0.728 Taiwan & 0.847 Mauritius \\ 
0.08 Netherlands & 0.703 Egypt & 0.843 Austria \\ 
0.069 Turkey & 0.69 Saudi Arabia & 0.838 Trinidad-Tobago \\ 
0.065 Sweden & 0.685 Mexico & 0.832 Sweden \\ 
0.062 South Africa & 0.682 Yugoslavia & 0.829 Cyprus \\ 
0.056 Austria & 0.68 Vietnam North & 0.827 Israel \\ 
    \hline
    \textbf{Group 4}     & \textbf{Group 5}  & \textbf{Group 6}       \\
    \hline
0.319 Yemen & 0.188 Djibouti & 0.849 Liechtenstein \\ 
0.296 Brunei & 0.187 Bhutan & 0.825 St Kitts-Nevis \\ 
0.276 Bahamas & 0.173 Guinea-Bissau & 0.775 Antigua-Barbuda \\ 
0.274 Singapore & 0.173 Swaziland & 0.747 Vanuatu \\ 
0.27 Cambodia & 0.155 Comoros & 0.737 Dominica \\ 
0.269 Angola & 0.151 Equatorial Guinea & 0.737 St Vincent-Grenadines \\ 
0.263 Senegal & 0.145 Qatar & 0.726 St Lucia \\ 
0.261 Mozambique & 0.145 Bahrain & 0.678 Western Samoa \\ 
0.257 Tanzania & 0.142 Gabon & 0.674 Grenada \\ 
0.253 Tunisia & 0.136 Cape Verde & 0.671 Seychelles \\ 
0.251 Namibia & 0.134 Malawi & 0.65 Belize \\ 
0.249 Afghanistan & 0.13 Oman & 0.644 Sao Tome Principe \\ 
0.248 Nepal & 0.125 Lesotho & 0.618 Maldives \\ 
0.244 Ghana & 0.122 St Kitts-Nevis & 0.516 Barbados \\ 
0.243 Kenya & 0.117 Mauritania & 0.515 Comoros \\ 
    \hline
  \end{tabular}
  \caption{\textbf{States with Highest Membership in Latent Groups,
      Cold War period}.  To identify the states with highest
    membership in each latent group, we average over each states'
    latent membership probabilities in the years 1950-1990.  Average
    group membership is reported beside the state name for the top 15
    states in each latent group.}
 \label{tb:CWstates}
\end{table}

\clearpage
\subsection{Estimated coefficients associated with hidden Markov state 2}

The main text focuses on effects during time periods associated with
hidden Markov state 1. For completeness, we present the results
associated with hidden Markov state 2.

\begin{table}[h] \spacingset{1} \setlength{\tabcolsep}{6pt} \centering
{\footnotesize
  \begin{tabular}{l......}  \toprule \textbf{Predictor} & \multicolumn{1}{c}{\textbf{Group
1}} & \multicolumn{1}{c}{\textbf{Group 2}} &
\multicolumn{1}{c}{\textbf{Group 3}} &
\multicolumn{1}{c}{\textbf{Group 4}} &
\multicolumn{1}{c}{\textbf{Group 5}} &
\multicolumn{1}{c}{\textbf{Group 6}} \\ \midrule {\tt INTERCEPT}  &
10.420 & 16.239 & 12.357 & 11.622 & 4.457 & 4.549 \\ &  (1.021) &
(1.021) & (1.021) & (1.021) & (1.060) & (1.064) \vspace{2mm} \\ {\tt
POLITY}  & -0.005 & -0.137 & 0.209 & 0.052 & -0.201 & -0.157 \\ & 
(0.914) & (0.913) & (0.913) & (0.913) & (1.047) & (1.062) \vspace{2mm}
\\ {\tt MILITARY} &  0.363 & 1.017 & 0.237 & 0.163 & -0.443 &
-0.556 \\ {\tt CAPABILITY}&  (1.063) & (1.062) & (1.062) & (1.061) &
(1.062) & (1.064) \\  \midrule \multicolumn{7}{l}{$N$ nodes: 216; $N$
dyad-years: $842,685$; $N$ time periods: 195}\\ \multicolumn{7}{l}{Lower
bound at convergence: $-527,587.7$}\\ \bottomrule
\end{tabular} }
\caption{\textbf{Estimated Coefficients and their Standard Errors, Markov State 2}.
The table shows the estimated coefficients (and standard errors) of
the two monadic predictors for each of six latent groups in the second Markov state.  The estimated coefficients for cubic splines and
indicators for variable missingness are not shown.}\label{tb:covFX_Markov2}
\end{table}

\clearpage
\subsection{Results of model estimated using online estimation approach}

The main text uses the entire set of observed networks, from 1816 to
2010, to estimate the model's parameters. As all time periods are
treated as observed at the time of analysis, this can be thought of as
performing \emph{batch} data analysis. An alternative approach
considers data streaming in in sequence --- either year by year or (as
we do illustrate here) after a mini-batch of years has been
observed. To illustrate this approach, we first fit a model for the
years 1816--1820, then use the resulting mixed membership estimates as
starting values for the model estimated by adding the next window (1821-1825).
We repeat until all years are included.

\begin{figure}[h!t] \spacingset{1}
     \includegraphics[scale=.6]{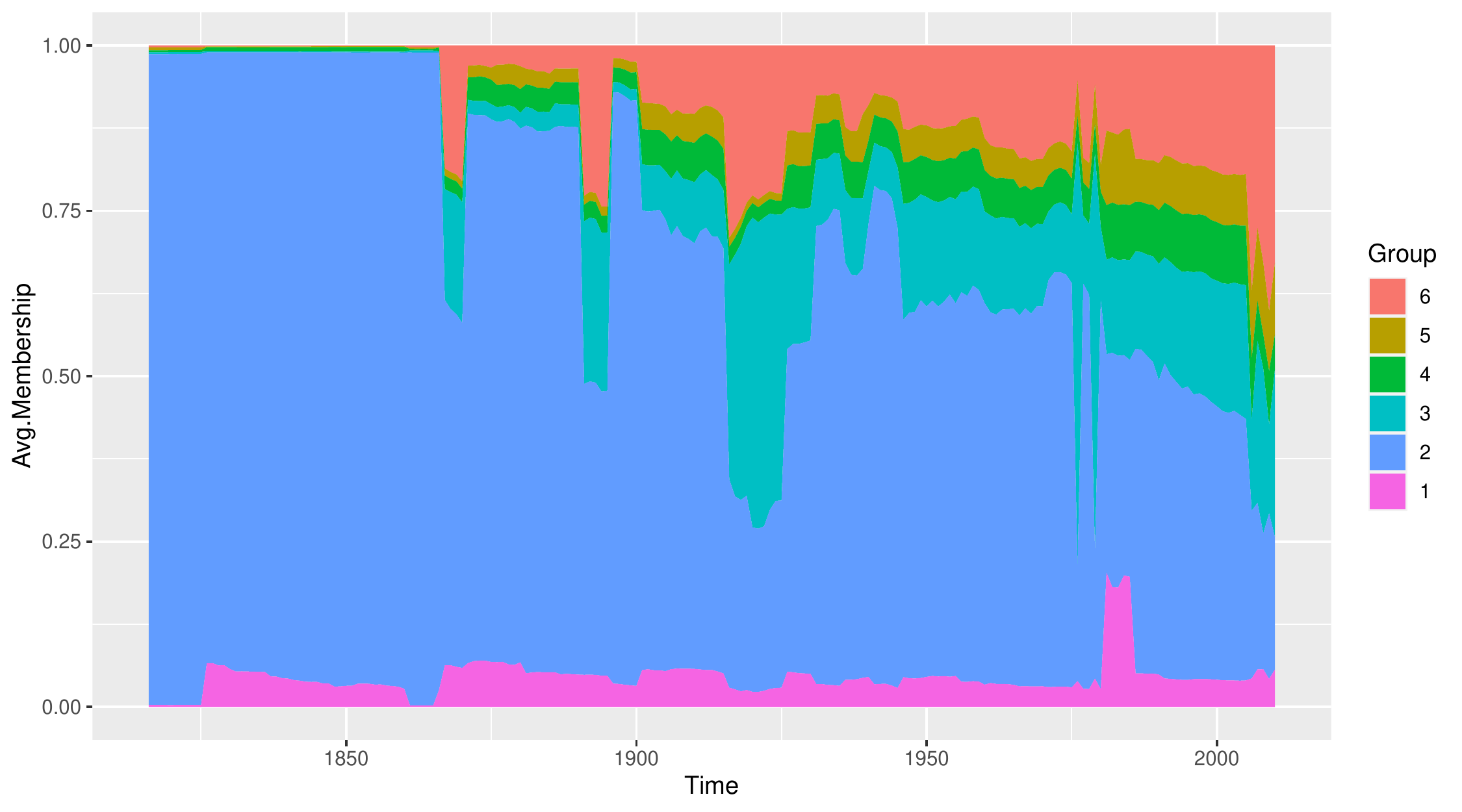} 
\vspace{.1in}
\caption[cluster time]{\textbf{Membership in Latent Groups over Time
    (online update model)}. The figure shows the average proportion of
  membership in six latent groups for each year from 1816--2010.
  Estimates are derived from specifications using expanding five-year
  windows.}
  \label{fig:cluster_time_online}
\end{figure}

\begin{figure}[h!t] \spacingset{1}
     \includegraphics[scale=.52]{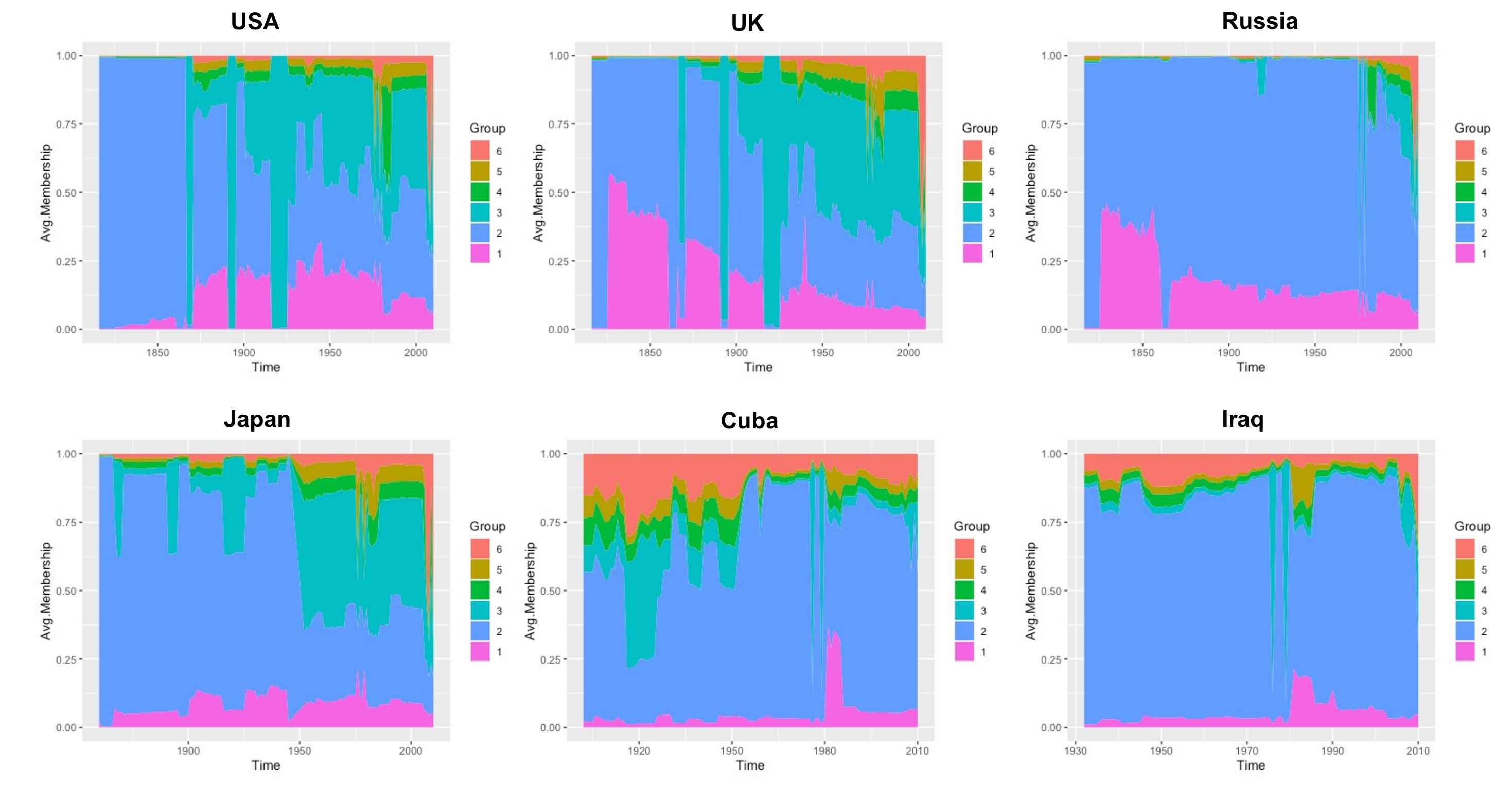} 
\vspace{-.1in}
\caption[cluster time node]{\textbf{Average Node Membership over Time,
Select States (online update model)}. The figure shows, for six states, the average rate of
membership in six latent groups in each year the state is present in
the network.  Estimates are derived from specifications using expanding five-year windows. }
  \label{fig:cluster_time_node_online}
\end{figure}

Figure~\ref{fig:cluster_time_online} we show how estimated memberships
evolve over the entire period under study using this online estimation
strategy. Membership patterns are positively correlated with the model
reported in the main text (0.516), but the evolution of membership
differs in several ways.  Group~2 is significantly larger throughout
the period, and the late increase in Group~6 (beginning in 2005) is more
pronounced. Figure~\ref{fig:cluster_time_node_online} shows the same
evolution, disaggregated by some of the countries we used in the main
text. Many previously observed structural breaks are apparent in these estimates
(e.g., Russia at the end of the Cold War, Japan in 1945), while others are attenuated (Cuba in the 1950s) or absent (Iraq in 1991). Finally,
Table~\ref{tb:covFX_online} reports estimated coefficient (and
associated standard errors) for the final time window in the online
estimation approach (viz. years 2006-2010).

\begin{table}[t] \spacingset{1} \setlength{\tabcolsep}{6pt} \centering
{\footnotesize
  \begin{tabular}{llllllll}  \toprule \textbf{Predictor} &
\multicolumn{1}{c}{\textbf{Dyadic}} & \multicolumn{1}{c}{\textbf{Group
1}} & \multicolumn{1}{c}{\textbf{Group 2}} &
\multicolumn{1}{c}{\textbf{Group 3}} &
\multicolumn{1}{c}{\textbf{Group 4}} &
\multicolumn{1}{c}{\textbf{Group 5}} &
\multicolumn{1}{c}{\textbf{Group 6}} \\ \midrule {\tt INTERCEPT} & &
$-0.997$ & $1.747$ & $-1.354$ &  $-1.817$ & $-2.602$ & $-2.384$  \\ 
& & (0.101) & (0.101) & (0.100) & (0.102) & (0.102) & (0107) \vspace{2mm}\\ 
{\tt POLITY} & & $-0.009$ & $-0.081$ &  $0.104$  & $0.024$ &  $0.027$ & $0.063$ \\ 
& & (0.102) & (0.101) & (0.089) & (0.105) & (0.110) & (0.151) \vspace{2mm}
\\ {\tt MILITARY} & & $-0.036$ &  $0.178$ & $-0.170$ &  $-0.171$ &  $-0.273$ &  $-0.332$ \\ {\tt CAPABILITY}& & (0.103) & (0.101) & (0.109) & (0.106) &(0.101) & (0.123) 
\\ \midrule {\tt BORDERS} & 2.374 &&&&&& \\ & (0.003)
&&&&&&\vspace{2mm} \\ {\tt DISTANCE} & -0.0001 &&&&&& \\ & (0.000)
&&&&&& \vspace{2mm} \\ {\tt ALLIANCE} & 0.082 &&&&&& \\ & (0.003)
&&&&&& \vspace{2mm} \\ {\tt IO CO-MEMBERS} & 0.002 &&&&&& \\ & (0.000)
 &&&&&& \vspace{2mm} \\ {\tt PEACE YRS} & -0.019 &&&&&& \\ & (0.000)
 &&&&&&\\ \midrule \multicolumn{8}{l}{$N$ nodes: 216; $N$
dyad-years: $842,685$; $N$ time periods: 195}\\ \multicolumn{8}{l}{Lower
bound at convergence: $-527,686.5$}\\ \bottomrule
\end{tabular} }
\caption{\textbf{Estimated Coefficients and their Standard Errors (online update model)}.
The table shows the estimated coefficients (and standard errors) of
the two monadic predictors for each of six latent groups, as well as
those of the dyadic predictors for edge formation.   Estimates are derived from specifications using expanding five-year windows. We report coefficient estimates for the final model (2006-2010).}\label{tb:covFX_online}
\end{table}

\clearpage

\subsection{Effects of decreasing POLITY}
In the main text, we focus on the effects of increasing POLITY scores (i.e. making countries more democratic). The proportion of states with the maximum polity score generally increases over time (approximately 17.5\% of states in 2010), so the effects we see may be driven by a sort of celing effect, as these countries are as democratic as they can be.  To ensure the estimated effect of POLITY is not driven by a ceiling effect, we analyze the effects of decreasing POLITY scores.  The results are substantively identical to those reported in the main text, mirroring the pattern shown in Figures~\ref{fig:CountryHet2} and~\ref{fig:politytime}.

\begin{figure}[h!t] \spacingset{1}
  \begin{center} \includegraphics[scale=.615]{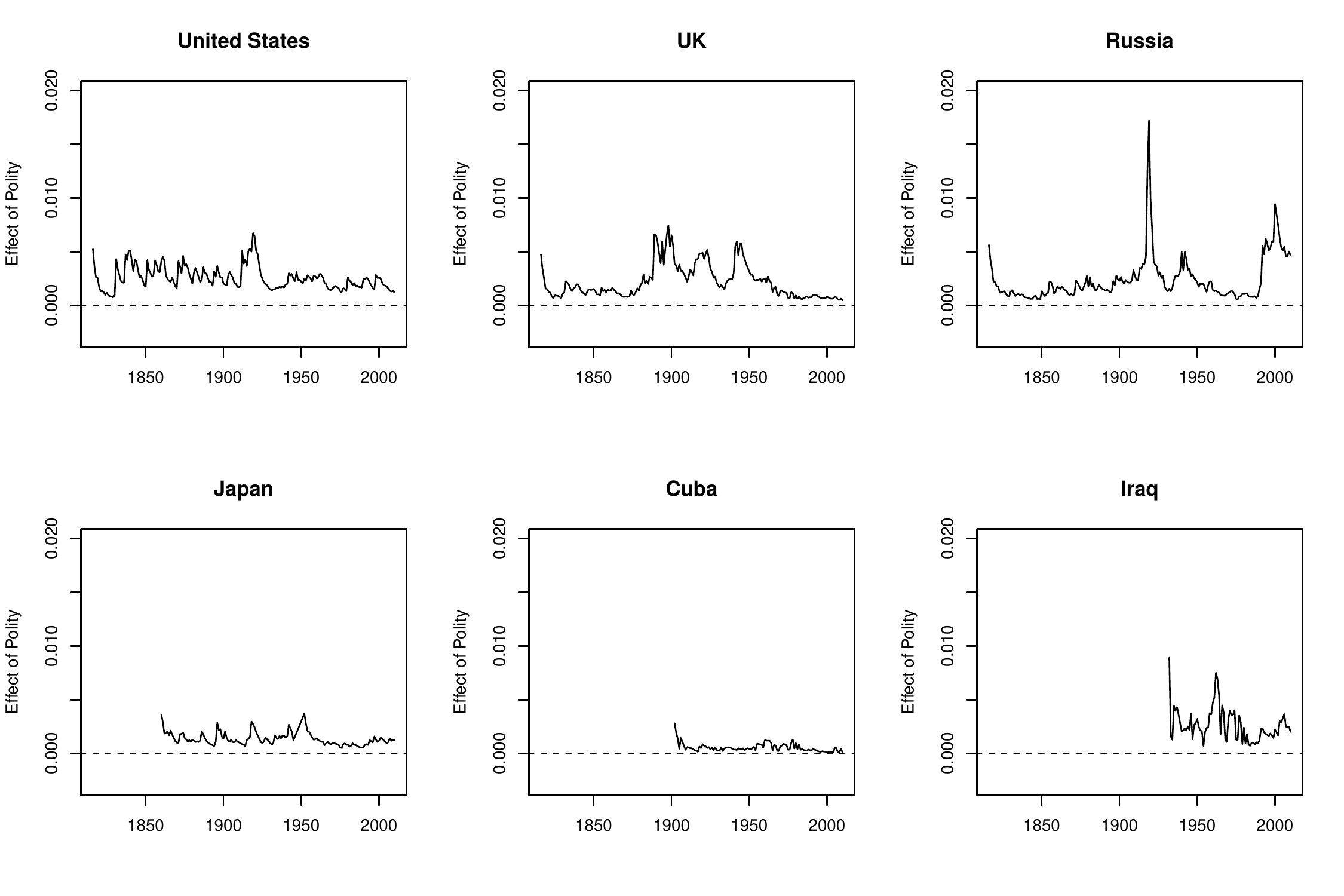}
 \end{center} \vspace{-.3in}
 \caption[CountryHet2_decrease]{\textbf{Effect of Decrease in Polity over Time,
Select States}.  The figure shows the estimated change in the
probability of interstate conflict over time if a country's {\tt POLITY}
score is decreased by one standard deviation (6.78) from its observed
value (down to a minimum of -10).}\label{fig:CountryHet2_decrease}
\end{figure}

\begin{figure}[h!t] \spacingset{1}
  \begin{center} \includegraphics[scale=.65]{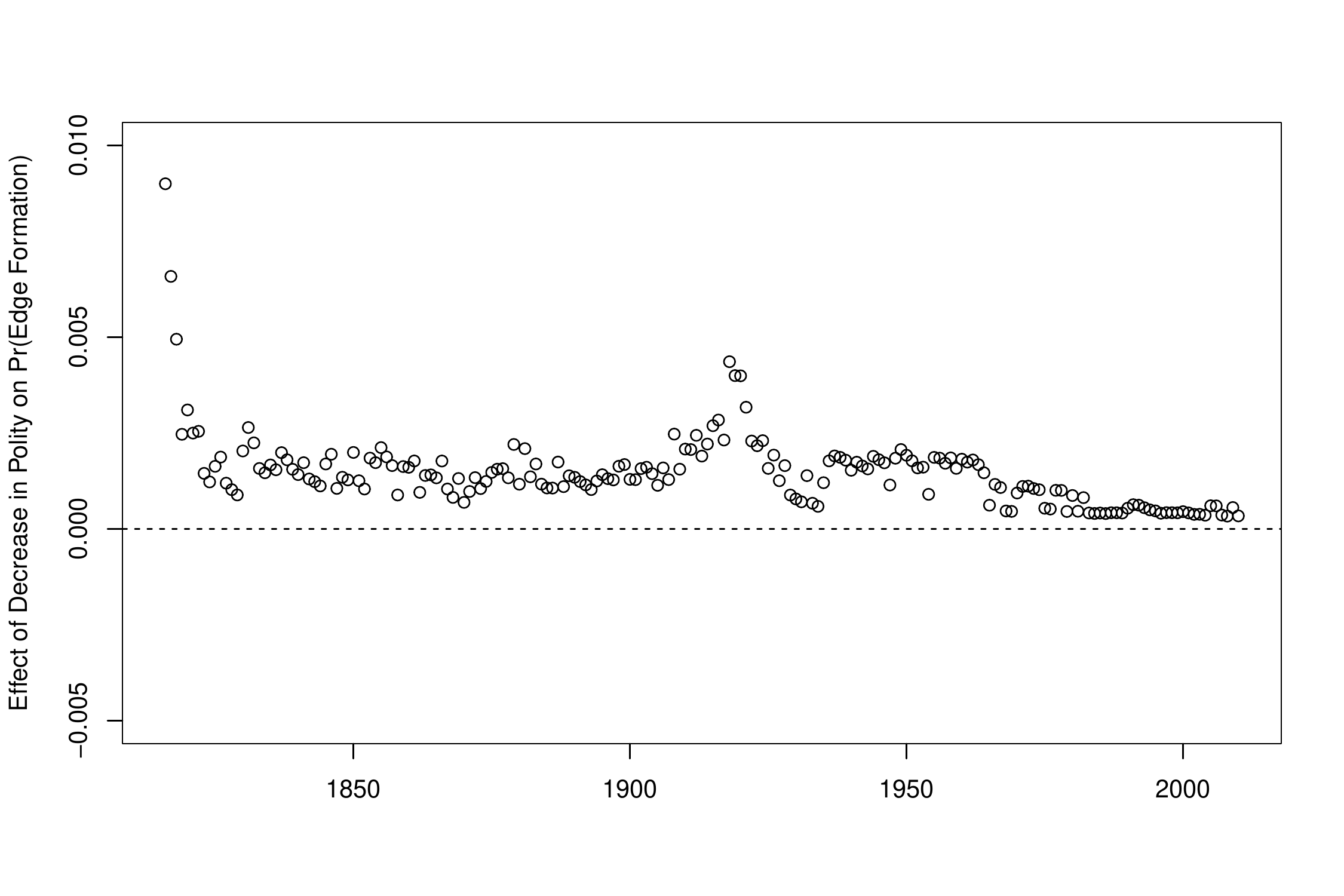}
 \end{center} \vspace{-.5in}
 \caption[politytime_decrease]{\textbf{Estimated Aggregate Effect of Decrease in
Polity over Time}.  The figure shows the estimated average change in
the probability of interstate conflict when countries' {\tt POLITY}
scores are decreased by one standard deviation (6.78) down to the minimum {\tt POLITY} score.}\label{fig:politytime_decrease}
\end{figure}

\clearpage
\subsection{Comparison with Logistic Regression}
\label{app:logitcompare}

In this section, we compare the forecasting performance of the \dynMMSBM{} to
that of the standard logistic regression model prevalent in the
democratic peace literature.  We fit this regression model to the same
interstate conflict data organized in the dyad-year format using the
identical set of predictors.  The only difference is that, in keeping with
the convention in the literature, we transform the monadic variables
({\tt POLITY} and {\tt MILITARY CAPABILITY}) to a dyadic structure.  We
follow the conventional approach to specifying {\tt POLITY}
 by including two separate variables measuring the democracy level of
 the less democratic country and that of the more democratic country
 in a dyad \citep[e.g.,][]{
dafoe2013democratic}.  The
 {\tt MILITARY CAPABILITY} variable is restructured as the ratio of the more
powerful state's military capability to the less powerful state's
military capability.

We then conduct an out-of-sample forecasting exercise on the years
2009-2010, which were excluded from our initial sample.  We follow
\citet{goldstone2010global} in using a 2-year window for out-of-sample
validation.  We use the parameters of the \dynMMSBM{} and logit models
to predict the onset of conflict for dyad-years in the 2009--2010
period.  Because the models include peace years and cubic splines as
predictors, we impute these variables based on estimated probabilities
of conflict in the out-of-sample set.  To impute, we first forecast
conflict in the year 2009 and then sample from the predicted
probabilities of conflict to update the peace years variable for each
dyad.  For the \dynMMSBM{}, we let the network evolve according to the
estimated Markov transition probabilities.

We evaluate the predictive accuracy of both models by comparing their
predictions with the observed pattern of conflict in 2009--2010.
First, we conduct a Diebold-Mariano test of comparative forecasting
accuracy \citep{Diebold1995, Harvey1997}.  The test, which compares
mean-squared forecasting error of the two methods, indicates that the
\dynMMSBM{} significantly outperforms the logit model in dispute
prediction (DM statistic = $-2.12$, $p$ = $0.034$).

\begin{figure}[ht] \spacingset{1}
  \centering
  \includegraphics[scale=0.5]{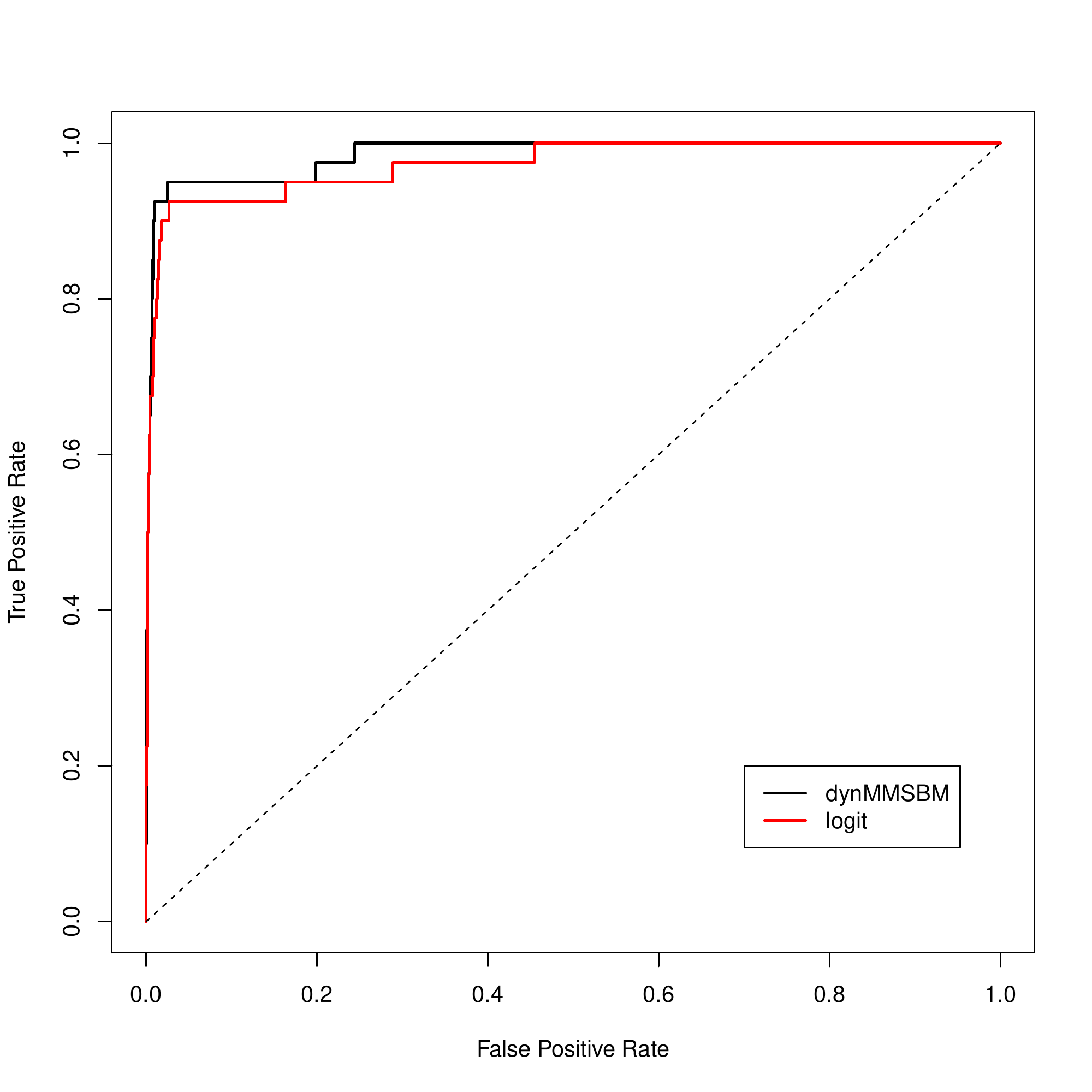}
  \caption[Forecast]{\textbf{ROC Curve: Logit, Dynamic
      Mixed-membership SBM Models}.  To perform the forecast, we
    exclude the final two years (2009-2010) from the dataset and
    estimate each model on the preceding years (1816-2008). Then we
    predict the missing years based solely on the covariate data.  }
\label{fig:Forecast}
\end{figure}

Second, we compare the receiver operating characteristic curves (ROCs)
for each model.  We display the ROC curves in
Figure~\ref{fig:Forecast} and show the area under the ROC curves in
\ref{tb:logit_AUC}.  By this criterion, the \dynMMSBM{} continues to
outperform the logit model but only marginally.  The \dynMMSBM{} has a
larger area under the ROC curve, though the difference is not
statistically significant.

\begin{table}[ht]
\spacingset{1}
\centering
\begin{tabular}{ll}
  \hline
Model & AUROC  \\ 
  \hline \\
\dynMMSBM{} & 0.986  \\ 
 & (0.013)  \\ \\
Logit & 0.973  \\ 
   & (0.018)  \\ 
   \hline
\end{tabular}   \spacingset{1}
\caption{\textbf{Out of Sample Prediction, \dynMMSBM{} vs. Logit}. The
  table displays the area under the ROC curve (AUROC) and associated
  standard error for the two models.  Each model is fit on data from
  1816-2008 and used to forecast conflict in the period
  2009-2010.} \label{tb:logit_AUC}
\end{table}

\bigskip
\pdfbookmark[1]{References}{References}
\spacingset{1.75}
\printbibliography

\end{document}